%% file: main.tex
\documentclass[sigconf]{acmart}

\makeatletter
\renewcommand{\@secnumfont}{}
\renewcommand\abstractname{\uppercase{Abstract}}
\makeatother

\usepackage{enumitem,pifont,xcolor,booktabs,array,multirow}
\usepackage{subcaption}
\usepackage{colortbl}

\newcommand{\sname}{\emph{\textit{SonicSieve}}}
\newcommand{\sssec}[1]{\vspace*{0.05in}\noindent\textbf{#1}}

\newcommand{\kuang}[1]{\textcolor{black}{#1}}
\newcommand{\swarun}[1]{\textcolor{black}{#1}}
\newcommand{\rr}[1]{\textcolor{black}{#1}}

\newcolumntype{L}[1]{>{\RaggedRight\arraybackslash}p{#1}}
\newcolumntype{C}[1]{>{\centering\arraybackslash}p{#1}}
\newcommand{\cmark}{\textcolor{green!60!black}{\ding{51}}}  
\newcommand{\xmark}{\textcolor{red}{\ding{55}}}             

\newcommand{\squishlist}{\begin{itemize}[itemsep=1pt,parsep=2pt,topsep=3pt,partopsep=0pt,leftmargin=0em, itemindent=1em,labelwidth=1em,labelsep=0.5em]}
\newcommand{\squishend}{\end{itemize}}
\newcommand{\squishenum}{\begin{enumerate}[itemsep=1pt,parsep=2pt,topsep=3pt,partopsep=0pt,leftmargin=0em, itemindent=1.5em,labelwidth=1em,labelsep=0.5em]}
\newcommand{\squishsubenum}{\begin{enumerate}[itemsep=1pt,parsep=2pt,topsep=0pt,partopsep=0pt,leftmargin=0em,listparindent=1.5em,labelwidth=1em,labelsep=0.5em]}
\newcommand{\squishenumend}{\end{enumerate}}

\AtBeginDocument{%
  }

\copyrightyear{2026}
\acmYear{2026}
\setcopyright{cc}
\setcctype{by}
\acmConference[CHI '26]{Proceedings of the 2026 CHI Conference on Human Factors in Computing Systems}{April 13--17, 2026}{Barcelona, Spain}
\acmBooktitle{Proceedings of the 2026 CHI Conference on Human Factors in Computing Systems (CHI '26), April 13--17, 2026, Barcelona, Spain}
\acmPrice{}
\acmDOI{10.1145/3772318.3790376}
\acmISBN{979-8-4007-2278-3/2026/04}

\begin{document}

\author{Kuang Yuan}
\authornote{Co-primary authors}
\affiliation{
  Carnegie Mellon University
  \country{USA}
}
\email{kuangy@andrew.cmu.edu}

\author{Yifeng Wang}
\authornotemark[1]
\affiliation{
  Carnegie Mellon University
  \country{USA}
}
\email{yifengw3@andrew.cmu.edu}

\author{Xiyuxing Zhang}
\authornotemark[1]
\affiliation{
  Tsinghua University
  \country{China}
}
\email{zxyx22@mails.tsinghua.edu.cn}

\author{Chengyi Shen}
\affiliation{
  Zhejiang University
  \country{China}
}
\email{shenchengyi@zju.edu.cn}

\author{Swarun Kumar}
\affiliation{
  Carnegie Mellon University
  \country{USA}
}
\email{swarun@cmu.edu}

\author{Justin Chan}
\affiliation{
  Carnegie Mellon University
  \country{USA}
}
\email{justinchan@cmu.edu}







\title{SonicSieve: Bringing Directional Speech Extraction to Smartphones Using Acoustic Microstructures}

\renewcommand{\shortauthors}{Yuan et al.}

\input{abs-4}



\begin{teaserfigure}
\centering
  \includegraphics[width=\textwidth]{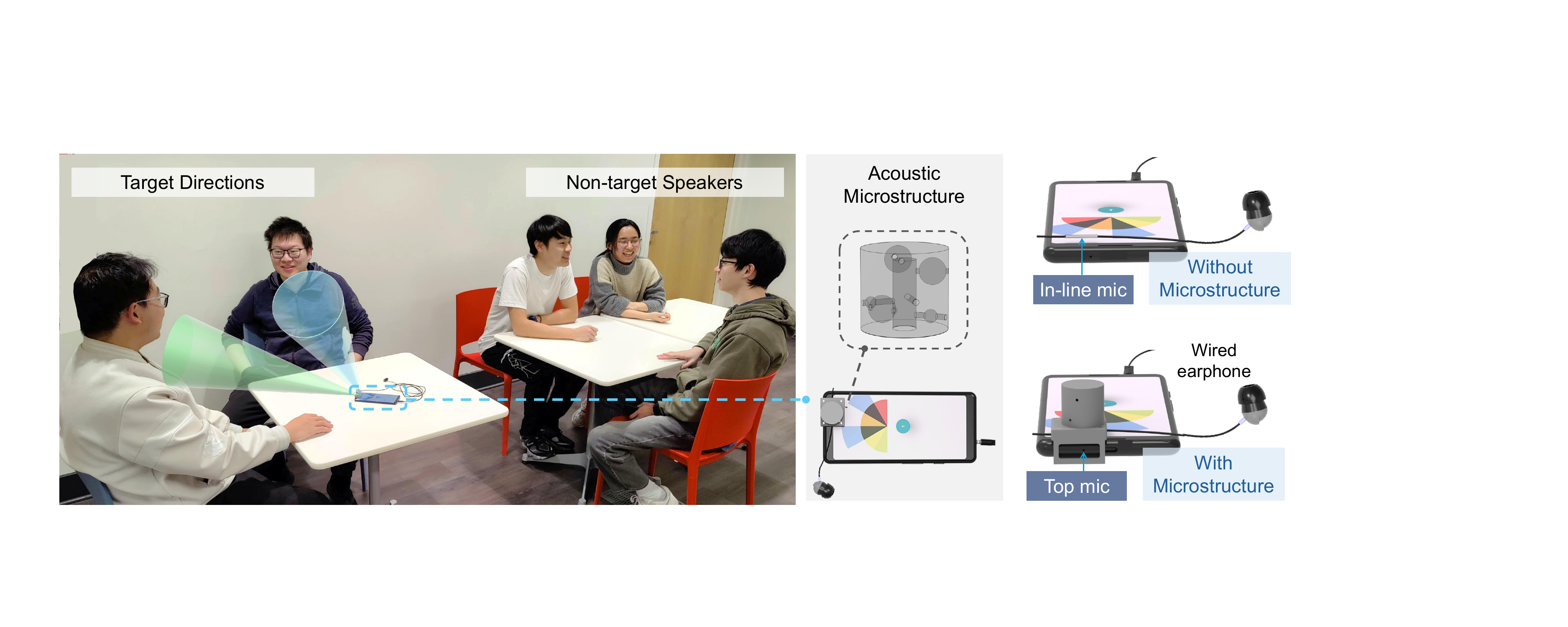}
  \vspace{-1.7em}
  \caption{{\bf {\sname} enables directional speech extraction on smartphones using a lightweight, passive acoustic microstructure.} {\bf (Left)} Our system leverages the distinct spatial cues created by the 3D-printed microstructure with a real-time neural network to intelligently amplify speech from target directions while attenuating others. {\bf (Right)} Our design attaches to the in-line microphone of low-cost earphones which can be plugged into a smartphone. The system records sound mixtures from the in-line and top microphone which are used by the neural network to generalize across different sound sources and environments.}
  \vspace{1em}
  \label{fig:concept}
  \Description{Three sub-figures illustrate how the system can assist with directional speech extraction in a meeting scenario. The left sub-figure displays a meeting room with two tables next to each other. At one of the tables are two of the target speakers having a discussion, there is a smartphone with the microstructure on the table. At the other table is a group of three non-target speakers producing interfering speech. Green and blue cones emanate from the smartphone and the microstructure pointing towards the target speakers illustrating the directional sound focus of SonicSieve. The middle sub-figure showcases the structure of the acoustic microstructure and the smarpthone user interface. The acoustic microstructure consists of a two-cylinder design with an outer cylinder containing holes that connect to the inner cylinder through tubes of varying lengths. These tubes feature differently sized spherical resonators attached along their length. The user interface displays 6 colored uniform sectors (30° each) forming a semicircle, indicating the selected direction with darkened color. The rightmost sub-figure shows how the microstructure is installed on the smartphone. The top part illustrates how the in-line microphone of a low-cost earphone is positioned on the smartphone, next to the smartphone's built-in top mic. The bottom part shows the acoustic microstructure mounted on top of the in-line mic of the earphone using a 3D-printed holder that couples together the microstructure and earphone to the smartphone.}
\end{teaserfigure}

\maketitle

\input{intro-8}
\input{related-3}

\input{system-3}
\input{network-2}

\input{eval-2}

\input{benchmark-2}
\input{ui-2}
\input{discussion-1}

\bibliographystyle{ACM-Reference-Format}
\bibliography{bib}
\end{document}

%% file: abs-4.tex
\begin{abstract}
Imagine placing your smartphone on a table in a noisy restaurant and clearly capturing the voices of friends seated around you, or recording a lecturer’s voice with clarity in a reverberant auditorium. We introduce {\sname}, the first intelligent directional speech extraction system for smartphones using a bio-inspired acoustic microstructure. Our passive design embeds directional cues onto incoming speech without any additional electronics. It attaches to the in-line mic of low-cost wired earphones which can be attached to smartphones. We present an end-to-end neural network that processes the raw audio mixtures in real-time on mobile devices. Our results show that \sname\ achieves a signal quality improvement of 5.0~dB when focusing on a 30\textdegree\  angular region. Additionally, the performance of our system based on only two microphones exceeds that of conventional 5-microphone arrays.
\end{abstract}

%% file: intro-8.tex
\section{INTRODUCTION}

In this paper, we ask the question: \textit{Can we enable directional speech extraction on smartphones?} A positive answer can enable smartphones to intelligently amplify target speakers and suppress unwanted interference. Imagine having a meeting in a noisy restaurant, and you only want to capture the speech of colleagues at your table for clear meeting notes and to remotely stream to other colleagues back at the office (Fig.~\ref{fig:concept}). Or imagine being a student in a classroom and you want to record the sound of the lecturer without needing a wearable microphone or a dedicated setup.

While rich prior work~\cite{pandey2024all,kovalyov2023dsenet,tesch2022insights,yang2024binaural,gu2024rezero,wechsler2023multimic,chen2024hearable,shen2020voice,wang2021using,wang2019contactless} on intelligent sensing for audio and speech using spatial information exist, they usually rely on arrays of microphones and estimate spatial cues of sounds by analyzing differences in arrival time and sound intensity across the microphones. Although microphone arrays have proliferated across various computing devices from smart speakers and smart glasses to AR/VR headsets, smartphones often contain only two microphones and remain largely unaware of the spatial richness of sound. Additionally, smartphones do not natively support external microphone arrays. While directional microphones exist, they only filter sound from a single, fixed direction, and don't provide multi-directional separation. 

We present {\sname}, the first intelligent directional speech extraction system on smartphones that uses a bio-inspired acoustic microstructure to embed distinct directional cues into incoming speech sounds from different angles.  
\kuang{The design of the acoustic microstructure} is inspired by directional hearing mechanisms found in nature. While human ears function like a two-microphone array, they are able to achieve directional hearing by leveraging spatial cues from the interaction of sound with the structure of the head and pinna~\cite{Brughera694356}. 
\kuang{\swarun{Perhaps closest to \sname\ in prior art }is Owlet~\cite{garg2021owlet}, a lightweight, miniature 3D-printed cylinder patterned with coded holes to encode spatial information into the signals arriving from different directions, enabling Direction-of-Arrival~(DoA) estimation for a single sound source.}
\kuang{\swarun{Inspired by Owlet}, \sname\ address a real-world user-centered challenge: going beyond localizing \textit{where} a sound signal originates, \sname\ tackles directional speech extraction that determines \textit{what} to extract from an acoustic mixture in a noisy multi-speaker environment.}

\kuang{To this end, we co-design both \swarun{novel} hardware and software \swarun{solutions for \sname}. \textit{First,} we optimize the dimensions and the coded pattern of the microstructure specifically for speech frequencies. We then create a practical design that attaches the microstructure to the in-line microphone of low-cost wired earbuds, which can be plugged into a smartphone.}
The system captures audio mixtures using both the in-line microphone and the smartphone’s co-located built-in top microphone, which serves as a reference (Fig.~\ref{fig:concept}) and enables the system to function reliably across diverse sound sources.
\kuang{\textit{Second,} we develop an end-to-end neural network that runs in real-time on the smartphone, processing the audio from both microphones to interpret the rich spatial cues and separate the speech from the target directions. }
We also design a smartphone user interface that lets users divide the surrounding space into a semicircular region with six 30° sectors. Users can select one or multiple sectors simultaneously, enabling functionality in multi-speaker scenarios (Fig.~\ref{fig:apps}) such as transcribing presentations with multiple speakers or conducting remote meetings in noisy environments like a park.

We build our end-to-end system on a smartphone with a low-cost wired earphone and a 3D-printed microstructure. 
\rr{This prototype therefore targets smartphones that can physically connect to a wired earphone and currently relies on an app-guided alignment procedure between the microstructure and the phone’s top microphone. We revisit these design trade-offs in our discussion section.} 
Our results are as follows: 
\squishlist
\item Our system achieves a 5.0~dB Scale-Invariant Signal-to-Distortion Ratio~(SI-SDR) improvement when focusing on a 30\textdegree\ sector for directional speech extraction, which significantly outperforms the baseline system without microstructure.
\item Our system demonstrates generalizable performance across 9 locations in 3 rooms.
\item Our average model inference time to process an 8~ms audio chunk is 7.12~ms and 4.46~ms on the Motorola Edge and Google Pixel 7 respectively, demonstrating real-time processing capability. 
\item Our user study with 20 participants rating 720 audio clips in total shows our system achieve a higher mean opinion score than a system based on a 5-channel microphone array.
\squishend

We will open source our microstructure design, code, and datasets, which can democratize directional speech extraction capabilities and make it available to the public. 

%% file: related-3.tex
\section{RELATED WORK}



\begin{table*}[t]
\small
\begin{tabular}
{|p{3cm}|p{1.2cm}|p{1.2cm}|p{2.2cm}|p{2.2cm}|p{2.2cm}|p{2.2cm}|}
\hline
\textbf{Reference} &
\centering\textbf{Mics} &
\centering\textbf{Single device} &
\centering\textbf{Directional speech extraction} &
\centering\textbf{Smartphone compatible} &
\centering\textbf{Real-world evaluation} &
\centering\arraybackslash\textbf{Real-time} \\
\hline
Conventional mic array & \centering 4--8+ & \centering \cmark & \centering \cmark & \centering \xmark & \centering N/A & \centering\arraybackslash N/A\\
\hline
Dia (2014)~\cite{sur2014autodirective} & \centering 2--8 & \centering \xmark & \centering \xmark & \centering \cmark  & \centering \cmark & \centering\arraybackslash \xmark\\
\hline
Owlet (2021)~\cite{garg2021owlet} & \centering 2 & \centering \cmark & \centering \xmark & \centering \xmark  & \centering \cmark & \centering\arraybackslash \cmark\\
\hline
DSENet (2023)~\cite{kovalyov2023dsenet} & \centering 3 & \centering \cmark & \centering \cmark & \centering \cmark  & \centering \xmark & \centering\arraybackslash \cmark\\
\hline
EarCase (2023)~\cite{li2023earcase} & \centering 2 & \centering \cmark & \centering \xmark & \centering \cmark   & \centering \cmark & \centering\arraybackslash DNS\\
\hline
Pandey et al. (2024)~\cite{pandey2024all} & \centering 8 & \centering \cmark & \centering \cmark & \centering \xmark  & \centering \xmark & \centering\arraybackslash \cmark\\
\hline
\rowcolor{gray!10}
\textbf{{\sname} (ours)} & \centering\textbf{\emph{2}} & \centering\cmark & \centering\cmark & \centering\cmark   & \centering\cmark & \centering\arraybackslash\cmark\\
\hline
\end{tabular}
\caption{{\bf Comparison of {\sname} with related systems.} {\sname} enables directional speech extraction across six $30^\circ$ sectors on smartphones in real-time, with evaluation in real-world environments. (DNS = did not specify)}
\end{table*}

To the best of our knowledge, no prior work has explored the use of acoustic microstructures to enable directional speech extraction on smartphones. Below we describe work related to acoustic microstructures, spatial sensing, and AI-enabled acoustic systems.

\noindent {\bf Acoustic sensing using microstructures.} Acoustic microstructures have been designed for a variety of applications, including spatial sensing~\cite{garg2021owlet,bai2022spidr}, sound absorption~\cite{arjunan2024acoustic,dupont2018microstructure}, acoustic filtering~\cite{li2016acoustic}, and biometric identification~\cite{yang2023biocase}. Owlet~\cite{garg2021owlet} designed a 3D-printed microstructure that embedded directional cues into incoming sounds, enabling direction-of-arrival (DoA) estimation and sound localization. SPiDR~\cite{bai2022spidr} extended this design by using microstructures to project and capture spatially coded signals to generate depth maps of nearby objects. 
EarCase~\cite{li2023earcase} adapted microstructures for smartphones, embedding them into a custom case for DoA estimation. However, its design required a new custom case for each device, limiting its ability to scale across multiple phone models and microphone layouts. \rr{In addition, case-based designs fix the relative placement between the microphones, which can make it harder to optimize for close microphone spacing that improves robustness across environments.} Our work differs in two key ways. First, our microstructure is designed for low-cost wired earphones, and is compatible with different smartphones. \rr{This form factor allows the in-line microphone to be positioned close to the phone’s built-in microphone, which empirically improves reliability under changing room acoustics (Sec.~\ref{subsec:integration}).} Second, our work is focused on directional speech extraction by leveraging spatial cues to selectively amplify and suppress speech based on their direction. This enables a broader set of spatial audio interactions beyond what is achievable through DoA alone.

\noindent {\bf Directional speech extraction.} Directional speech extraction is the task of isolating speech from a spatial region that is fixed or adjustable~\cite{pandey2024all}. Prior work~\cite{pandey2024all} explicitly relies on DoA information derived from 8-channel microphone arrays, and does not focus on smartphones. Related approaches~\cite{kovalyov2023dsenet,yang2024binaural} explore two- or three-microphone configurations found on some smartphones. However, the spatial cues provided by these configurations are fundamentally limited by both the small number of microphones and their suboptimal placement for spatial sensing tasks. In contrast to these works, we present an acoustic microstructure design which is optimized to enhance the diversity of spatial cues for speech signals. Our system does not require explicit DoA estimation, instead it takes the raw audio mixture as input, and uses an end-to-end network to implicitly infer spatial and acoustic cues, enabling real-time directional speech extraction on smartphones.

\noindent {\bf Beamforming using statistical algorithms.} Traditional beamforming techniques are typically based on statistical algorithms, including non-adaptive approaches such as Bartlett (delay-and-sum) and superdirective beamformers, as well as adaptive methods like minimum-variant distortionless-response (MVDR), and linearly constrained minimum variance (LCMV). \kuang{However, these algorithms are designed for conventional microphone arrays with fixed geometries}, and do not account for the complex direction-dependent transformations introduced by acoustic microstructures. As a result, they cannot be directly applied to microstructure-enabled systems like ours, where spatial cues arise not only from microphone placement but also from the physical design of the structure. Moreover, statistical beamformers are limited to capturing spatial cues, and lack the ability to model acoustic cues as neural beamformers do. 

\noindent {\bf Neural beamformers.} Prior work on neural beamformers have leveraged LSTM networks~\cite{chen2018multi,gu2019neural}, but are non-causal and are intended to support full-length audio inputs. Subsequent works~\cite{gu2020temporal,qian2018deep} introduced causal models that support online processing, but are not real-time. More recent work~\cite{veluri2024look,wang2022hybrid} has focused on designing causal real-time neural beamformers focused on directional hearing for AR headsets, smart glasses, and custom headphones equipped with four or more microphones that provide spatial cues. In contrast, we present a real-time neural network that can process the spatial and acoustic cues of audio that has been filtered by a microstructure, and use this for the task of directional speech extraction.


\noindent {\bf AI-enabled spatial-aware audio processing.} These systems have focused on speech enhancement~\cite{pascual2017segan,chatterjee2022clearbuds}, sound localization~\cite{shimada2021accdoa,nguyen2022salsa,yuan2025windDancer}, sound classification~\cite{Zhang2024EarSAVAS,Wang2024DreamCatcher,Zhang2023EarCough,yuan2024tomobrush} and speech separation~\cite{jenrungrot2020cone,wang2023tf}. Recent works have leveraged spatial information to enable new capabilities on smart devices. Cone of Silence~\cite{jenrungrot2020cone} extracts speech from multiple target speakers at different locations. ReZero~\cite{gu2024rezero} extracts all sounds within a user-defined spatial region—angular, spherical, or conical. \rr{However, most of these systems either require large microphone arrays with at least six elements, or are not designed for causal, real-time use.}
\rr{ClearBuds~\cite{chatterjee2022clearbuds} and Sound Bubble~\cite{chen2024hearable} achieve causal, real-time performance using wearable devices. ClearBuds leverages wireless earbuds~\cite{cao2020earphonetrack} for real-time speech enhancement in telephony, while Sound Bubble isolates audio within a defined radius. However, these systems are designed to enhance the wearer's hearing in human-human communication. In contrast, our work targets smartphone-based directional speech extraction for general audio recording and streaming applications, such as meeting transcription and voice assistants. Smartphones are well-suited for ad-hoc application scenarios as they are mobile devices that people carry throughout the day.}

\noindent {\bf Spatial sensing on smartphones.} Prior work has focused on estimating DoA~\cite{kuccuk2019real,zhou2024rethinking,tokgoz2020real}, and beamforming~\cite{gaubitch2014near} on smartphones using statistical algorithms and neural networks. However, the performance of these systems is fundamentally limited by the number of microphones. Ad hoc microphone arrays composed of smartphones have also been proposed for spatial sensing~\cite{sur2014autodirective}. These systems emulate larger arrays by synchronizing multiple devices. However, they rely on complex synchronization protocols such as aligning smartphone CPU clocks using WiFi beacon timestamps, synchronizing audio I/O clocks through a backend server, or using time-of-arrival techniques~\cite{li2017design,peng2007beepbeep}, which can make deployment in real-world scenarios challenging.

%% file: system-3.tex
\section{SONIC SIEVE}

This section outlines the technical approach of \sname.  \rr{Our goal is to let users \textit{pick a direction} on a phone interface and have the phone \textit{amplify that speaker} in real time, using a low-cost add-on rather than specialized microphone arrays.} We start with a high-level overview of how our system uses an acoustic microstructure for directional speech extraction. Then, we will detail the techniques for optimizing this microstructure, integrating it with a smartphone, using a neural network to extract speech in real-time, and real-world dataset collections.

\begin{figure}
    \centering
        \begin{subfigure}{0.48\linewidth}
        \includegraphics[width=\linewidth]{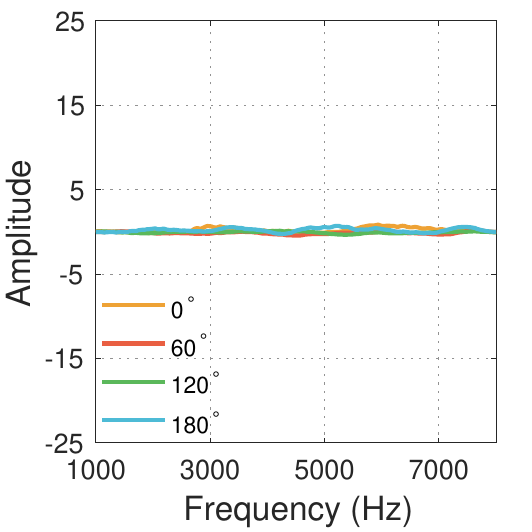}
        \caption{Without Microstructure}
    \end{subfigure}
    \hspace{-1mm}
    \begin{subfigure}{0.48\linewidth}
        \includegraphics[width=\linewidth]{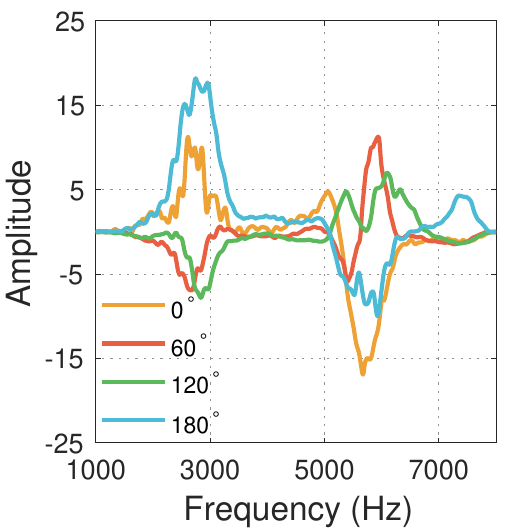}
        \caption{With Microstructure}
    \end{subfigure}
    \hspace{-1mm}

    \caption{{\bf Effect of microstructure on incoming sound signals across different angles of arrival.} The microstructure introduces larger variations in the frequency response $M_\theta(f)$ across different angles, providing enhanced spatial cues that can be leveraged for directional speech extraction.}
    \Description{This figure shows two frequency response graphs labeled "(a) Without Microstructure" and "(b) With Microstructure" displaying amplitude of frequency response $M_\theta(f)$ (y-axis, ranging from -25 to 25) versus frequency in Hz (x-axis, ranging from 1000 to 8000 Hz). In graph (a) Without Microstructure, four colored lines representing different direction of arrival (0°, 60°, 120°, and 180°) show nearly flat, overlapping frequency responses with minimal variation between angles, all hovering around 0 amplitude across the frequency spectrum. In contrast, graph (b) With Microstructure shows the same four angles producing dramatically different frequency response patterns. The lines show significant peaks and valleys, particularly around 3000 and 6000 Hz where the greatest variation occurs. The 0° line (orange) and 60° line (red) show pronounced peaks reaching amplitudes of approximately 15 and 10 respectively, while the 120° line (green) shows a deep trough reaching approximately -15 amplitude in the same frequency region. The 180° line (blue) exhibits multiple smaller peaks and valleys throughout the spectrum.}
    \label{fig:angle_diversity}
\end{figure}

\subsection{System Overview}
\label{sec:overview}

Traditional directional speech extraction relies on microphone arrays that use beamforming to isolate sound from specific directions. \rr{These arrays are common in dedicated devices (e.g., smart speakers), but are less accessible on everyday smartphones due to hardware constraints.} These systems work by analyzing timing and amplitude differences across multiple microphones—when sound arrives from a particular direction $\theta$, each microphone in the array receives the signal at slightly different times and intensities due to their spatial arrangement. The system can be modeled as:

\begin{equation}
    \mathbf{x}(f,t)=\mathbf{a}(f,\theta) s_{\theta}(f,t)+\mathbf{u}(f,t)
\end{equation}
where $\mathbf{a}(f,\theta)$ encodes the directional information based on known geometric delays between microphone channels, and $\mathbf{u}(f,t)$ represents interference and noise. \rr{Intuitively, $\mathbf{a}(f,\theta)$ describes how a sound from direction $\theta$ “shows up” across microphones, enabling spatial filtering.} Beamforming algorithms then apply spatial filtering to enhance signals from the target direction while suppressing others.

While various beamforming techniques work reliably for directional audio enhancement~\cite{adel2012beamforming, araki2007blind}, they typically requires a microphone array with multiple channels (usually 4-8) to handle the spatial ambiguity~\cite{Desai2022SSLReview, Francl2022DNN} and achieve satisfactory performance, which does not exist on smartphones. 

\begin{figure}
    \centering
    \includegraphics[width=0.8\linewidth]
{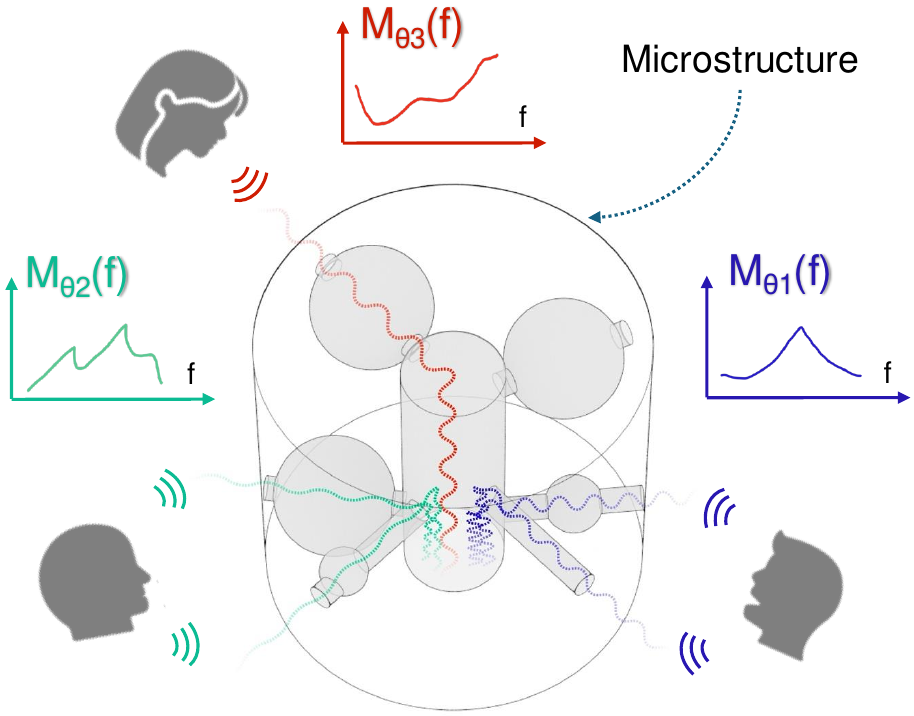}
\vspace{-0.5mm}
  \caption{{\bf Acoustic microstructure: principle of operation.} The structural elements of the microstructure (holes, tubes, and resonators) form a complex multipath environment that creates variations to incoming acoustic signals based on their direction of arrival.}
  \label{fig:stencil_concepts}
  \vspace{-0.5mm}
  \Description{This figure illustrates the operating principle of an acoustic microstructure through a conceptual diagram. At the center is a schematic representation of a microstructure, shown as a rounded enclosure containing multiple circular elements connected by pathways that processes sound differently based on direction. Three profile silhouettes of human heads are positioned around the microstructure: Top: A head emitting red sound waves (labeled with red frequency response graph Mθ3(f)). Left: A head emitting green sound waves (labeled with green frequency response graph Mθ2(f)). Right: A head emitting blue sound waves (labeled with blue frequency response graph Mθ1(f)). Each colored sound wave follows a different path through the microstructure, represented by matching colored wavy lines. The microstructure contains resonators and connecting tubes that create a complex multipath environment. Next to each head is a corresponding frequency response graph (Mθ1(f), Mθ2(f), and Mθ3(f)) showing how the microstructure transforms the acoustic signal from that particular direction, with each graph showing a distinctly different pattern: Red (top): A simple curve with a single peak. Green (left): A more complex pattern with multiple peaks and valleys. Blue (right): A smooth bell curve with a single prominent peak. }
\end{figure}

Instead of encoding the spatial information through microphone array geometry, \sname\ leverages a passive acoustic microstructure that directly embeds direction-dependent spatial cues into the signal. Such a design is inspired by nature's directional hearing system. For example, humans can localize and separated sound effectively with just two ears, which is due to the geometry of the head and ears creating distinctive reflecting and scattering effects that vary with sound angle~\cite{van2016auditory, xie2013head}. Similarly, owls achieve exceptional sound localization through asymmetric ear structures~\cite{coles1988directional_owl}. \rr{In practice, our microstructure acts like an acoustic lens placed over a microphone: it reshapes incoming sound differently depending on where it comes from (Fig.~\ref{fig:stencil_concepts}).}

Specifically, our acoustic microstructure applies a direction-specific filter $M_\theta(f)$ to a sound source arriving at the microstructure from a relative angle $\theta$ and propagating through the internal structure. We present an example in Fig.~\ref{fig:angle_diversity} of the direction-specific filter $M_\theta(f)$ at four different angles without and with the microstructure. \rr{The microstructure introduces substantially greater variation across different directions of arrival---without it, the signals look almost identical across angles; with it, each angle leaves a more distinctive spectral signature that a model can learn to exploit.}

The received signal at the microphone inside the microstructure can be denoted as:
\begin{equation} \label{eq:main_mic}
x(f,t)=M_{\theta}(f) s(f,t) + u(f,t)
\end{equation}

\rr{However, with only single-channel audio, we can not reliably isolate the spatial filter $M_{\theta}(f)$ in real-world settings. The challenge is that different sound sources have different intrinsic spectra, and room reflections further modify the signal. With one channel, the model cannot distinguish whether a spectral pattern comes from the microstructure's directional filtering, the source itself, or the room. A single-mic system might work for fixed sources in controlled environments, but cannot generalize across diverse real-world conditions.}


Thus, we incorporate a secondary reference microphone placed outside the microstructure. \rr{This reference gives the model a “before/after” view: one channel is shaped by the microstructure, and one is not. The reference microphone records the unmodified sound as:}
\begin{equation} \label{eq:ref_mic}
x_{ref}(f,t)=s'(f,t) + u_{ref}(f,t)
\end{equation}

We note that the received signal at the reference microphone from the target source $s'(f,t)$ differs slightly from the one at the inside microphone $s(f,t)$, since the two microphones are not positioned at the exact same location, such that the arriving signals are propagated through slightly different environmental reflections. \rr{To minimize this mismatch and enable robust estimation of spatial cues across different environments, we position the two microphones as close as practically feasible.}

Finally, we develop a deep neural network that processes the two-channel input to extract directional speech from the desired direction. Specifically, our directional speech extraction network $h$ estimates the target speech from the selected directions $\mathbf{\theta}$:
\begin{equation}
    \hat{s}_{\mathbf{\theta}}(f,t) = h\left(x(f,t), x_{ref}(f,t), \mathbf{\theta}\right)
\end{equation}
\rr{In other words, given the two recordings and the user-selected sector(s), the network outputs only the speech coming from those directions.}

\begin{figure}
    \centering
    \includegraphics[width=0.9\linewidth]{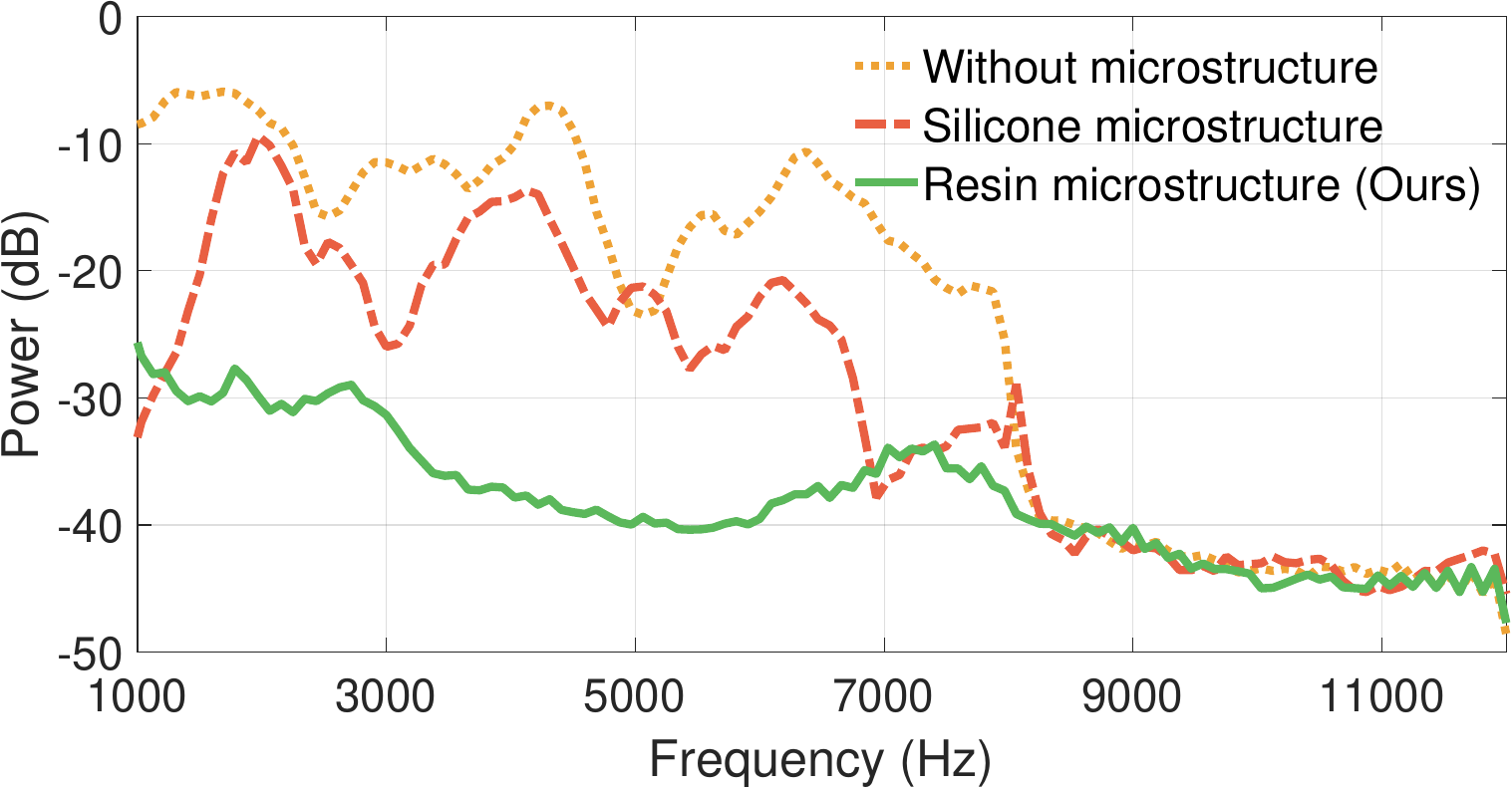}
  \caption{{\bf Effect of microstructure material (without holes) on acoustic attenuation.} The resin material effectively attenuates sound and ensures that it primarily travels through the microstructure’s holes, rather than its walls.}
  \label{fig:sound_isolation}
  \Description{This figure shows a line graph depicting the effect of different microstructure materials on acoustic attenuation. The graph plots power in decibels (dB) on the y-axis, ranging from 0 to -50 dB, versus frequency in Hertz (Hz) on the x-axis, ranging from 1000 to 12000 Hz. Three different conditions are represented by different colored lines: "Without microstructure" (orange dotted line at top): Shows the highest power levels, fluctuating between approximately -5 and -20 dB across most of the frequency range before dropping to around -45 dB at higher frequencies (8000-12000 Hz). "Silicone microstructure" (red line in middle): Displays moderate attenuation with power levels mostly between -10 and -30 dB, showing multiple peaks and valleys across the frequency spectrum before converging with the other conditions at higher frequencies. "Resin microstructure (Ours)" (green line at bottom): Demonstrates the strongest attenuation, particularly in the mid-frequency range (1000-8000 Hz) where it maintains a consistently lower power level around -30 to -40 dB.}
\end{figure}

To summarize, our system has four main components, which we describe below and present the details in the following sections:
\squishlist
\item \textit{Acoustic microstructure optimized for speech.} We present a compact acoustic microstructure that encodes spatial cues into the signal detectable by microphones. Building on prior microstructure work~\cite{garg2021owlet}, we further optimize the design specifically for spatial diversity at speech frequencies. 
\item \textit{Integration of acoustic microstructure to smartphones.} 
We present our design that can attach the acoustic structure to the in-line microphone of low-cost wired earbuds, which can be plugged into smartphones. This approach offers two key benefits: the in-line mic can be positioned close to the built-in mic to minimize signal differences, and the design can potentially be generalized to different smartphone models.
\item \textit{Real-time directional speech extraction network.}
While conventional beamformers work with known array geometries, they cannot handle the complex propagation patterns created by microstructures. We design a real-time neural network that learns the spatial and acoustic cues from our microstructure to extract speech from multiple selected directions simultaneously, enabling multi-speaker transcription for interviews and group discussions.
\item \textit{Real-world dataset.} Acoustic and spatial cues signatures learned in one environment may not generalize well to others, as variations in acoustic properties, such as reverberation~\cite{wu2016reverberation}, can distort the spatial cues encoded by the microstructure. To enable robust performance across different environments, we construct a diverse training dataset that captures the variability encountered in real-world acoustic environments. This dataset is designed to enable generalization across different spaces, speaker configurations, and background noise conditions.

\squishend

\subsection{Speech-aware acoustic microstructure}
\label{subsec:bg}

We build upon Owlet's microstructure design~\cite{garg2021owlet} which consists of a hollowed-out cylinder that encases a microphone.  The design incorporates three key structural elements: surface holes that act as virtual sound sources, capillary tubes with varying lengths that modify sound wave propagation, and Helmholtz resonators that selectively amplify different frequency bands. Together, these elements create unique acoustic signatures for sounds arriving from different angles.
As visualized in Fig.~\ref{fig:stencil_concepts}, when sound arrives from angle $\theta$, the microstructure applies a direction-specific filter $M_\theta(f)$. This directional fingerprint enables our system to distinguish between sounds from different locations. 

In this section, we present our approach to optimize the acoustic microstructure and detail the key design factors that impact the overall system performance below:

\begin{figure}
    \centering
    \vspace{-2mm}
    \includegraphics[width=0.78\linewidth]{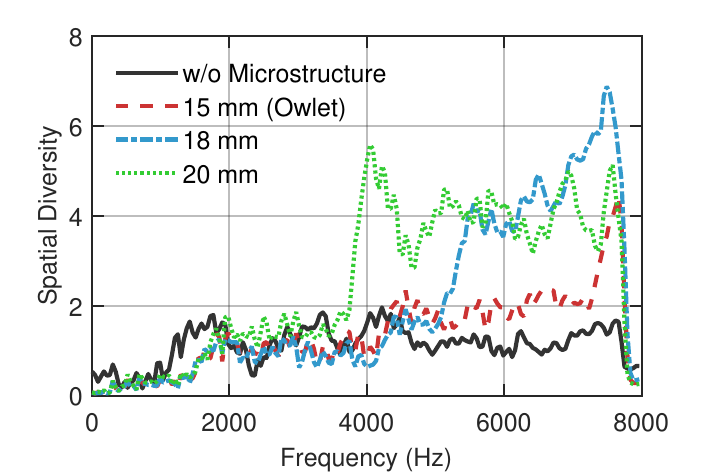}
     \vspace{-2mm}
    \caption{{\bf Effect of microstructure diameter on spatial diversity.} The microstructure with a larger diameter (20~mm) provides an overall higher spatial diversity across the speech frequencies.}
    \label{fig:freq_response_compare}
    \Description{The figure illustrates the effect of different microstructure diameters on spatial diversity across a range of frequencies. The x-axis is labeled "Frequency(Hz)" and ranges from 0 to 8000~Hz. The y-axis is labeled "Spatial Diversity" and ranges from 0 to 8. The figure contains four lines, each representing a different condition. The black solid line represents a condition "w/o Microstructure"(without microstructure). The red dashed line represents a 15~mm microstructure diameter, labeled as "15~mm(Owlet)". The blue dashed represents a 18~mm microstructure diameter. The green dotted line represents a 20 mm microstructure diameter. The plot shows that all four lines generally increase in spatial diversity as frequency increases, but their behavior varies. The "w/o Microstructure" line remains at the lowest spatial diversity across all frequencies. The 15 mm line shows a slight increase but remains relatively low. The 18 mm and 20 mm lines show significantly higher spatial diversity, with the 20 mm line generally outperforming the 18 mm line. The microstructure with a larger diameter provides an overall higher spatial diversity across the speech frequencies.}
\end{figure}

\begin{figure*}[t]
\centering
\includegraphics[keepaspectratio, height=0.166\textheight]
{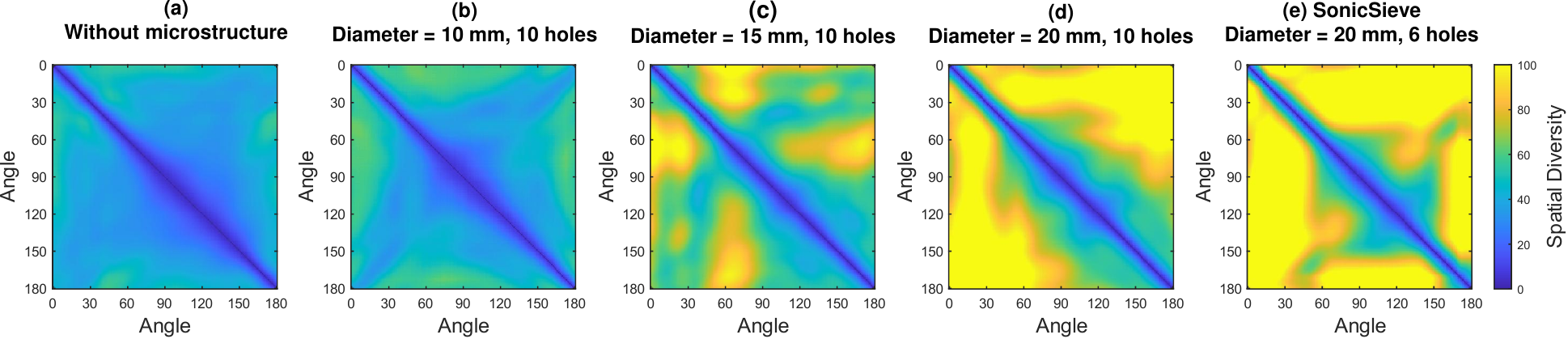}
  \vspace{-0.2em}
  \caption{{\bf A comparison of spatial diversity across different microstructure designs of varied diameter and hole number.} Spatial diversity is computed across angles in a semicircle.}
  \label{fig:diversity}
  \Description{This figure presents five heatmaps (a-e) comparing spatial diversity across different microstructure designs with varied microstructure diameters and hole numbers. Each heatmap displays "Angle" on both the x-axis and y-axis (ranging from 0 to 180 degrees), with color intensity representing "Spatial Diversity" according to the color bar on the right (ranging from 0 to 100, with dark blue indicating low diversity and bright yellow indicating high diversity). The five designs compared are: (a) Without microstructure: Shows predominantly low spatial diversity (dark blue). (b) Diameter = 10mm, 10 holes: Displays slightly increased diversity (light blue/green areas) compared to (a). (c) Diameter = 15mm, 10 holes: Shows significantly higher spatial diversity (light yellow/green areas) in several regions, particularly for the 45 to 75 degree. (d) Diameter = 20mm, 10 holes: Demonstrates even higher spatial diversity with extensive yellow regions covering most of the heatmap, indicating strong diversity across many angle combinations. Although there is no significant difference for the degree ranging from 120 to 180. (e) SonicSieve (Diameter = 20mm, 6 holes): Shows the highest overall spatial diversity with bright yellow dominating most of the heatmap. Compared to design of diameter = 20mm, 10 holes.}
\end{figure*}

\noindent \textbf{Material Selection for Acoustic Attenuation.} Our material selection aims to maximize sound attenuation, ensuring audio travels through the designed holes rather than leaking through the walls. We tested three configurations using a 1--8~kHz chirp at 70~dBA (normal conversation level): no microstructure, silicone microstructure, and resin microstructure.

As shown in Fig.~\ref{fig:sound_isolation}, the resin microstructure achieves 33~dB average sound reduction across tested frequencies, significantly outperforming silicone (17~dB reduction). This superior attenuation ensures sound primarily diffracts through the holes as intended, making resin our material of choice.


\noindent \textbf{Spatial Diversity Optimization.} To optimize performance for directional speech extraction, we maximize spatial diversity---the variation in how the structure responds to speech from different directions. Mathematically, we seek to maximize the variance $\mathbb{V}_\theta\left[|M_{\theta}(f)|\right]$, ensuring speech signals from different angles produce maximally different frequency responses. 
Speech signals are wideband acoustic signals with frequency components mainly under 8~kHz, and the components that are most critical for intelligibility are in 1-4~kHz~\cite{kunchur2023human}. We conduct experiments using wideband audio under 8~kHz in a constrained environment (anechoic chamber) on different microstructure designs.

To measure spatial diversity, we record speech signals from 0\textdegree\ to 180\textdegree\ at 1\textdegree\ resolution using speakers and microphones inside and outside the microstructure. Since the measurements are taken in an anechoic chamber, we can assume the environmental reflections and background noise are negligible (i.e. $s'(f,t)=s(f,t)$ and $u(f,t)=0$). From Eq.~\ref{eq:main_mic} and Eq.~\ref{eq:ref_mic}, the spatial diversity can then be calculated by:
\begin{equation}
    \mathbb{V}_\theta\left[|M_{\theta}(f)|\right] = \mathbb{V}_\theta\left[\frac{|x(f)|}{|x_{ref}(f)|}\right]
\end{equation}
where $x(f)$ and $x_{ref}(f)$ are signals from the internal and reference microphones. Based on the measurement setup, we optimize the following two key factors to improve the spatial diversity.

\squishlist
\item \textit{Microstructure Diameter.}  Fig.~\ref{fig:freq_response_compare} shows that Owlet's original 15~mm design achieves spatial diversity mainly at 7--8~kHz, with poor performance below 7~kHz, which can be potentially problematic for speech signals. Based on the principle that microstructure size correlates with resonating wavelengths, we conduct measurements on larger diameters. Both 18~mm and 20~mm diameters show improved diversity below 8~kHz, with 20~mm providing a better performance across speech frequencies.
We further visualize spatial diversity across different angles by calculating the pairwise spectral distance (i.e. $\left \lVert M_{\theta_1}(f) -  M_{\theta_2}(f) \right \rVert_2$) for the microstructure with diameters of 10~mm, 15~mm, and 20~mm. As shown in Fig.~\ref{fig:diversity} (a-d), the microstructure with a higher diameter achieves higher spatial diversity across a wider range of angles. We use 20~mm as our final design. We note that we did not further increase the diameter as a larger microstructure may obstruct the smartphone screen and hinder usability.


\item \textit{Number of Surface Holes.} As we observed in Fig.~\ref{fig:diversity}(d), the spatial diversity across the angles is not evenly distributed---angles $>60$\textdegree\ show lower diversity than 0--60\textdegree. To address this, we further optimized the surface hole pattern.  We empirically find that, too few holes limit directional sensitivity (sounds from non-hole directions get attenuated), while too many holes eliminate spatial diversity entirely. Thus, we selectively reduce the original ten-hole design to six holes by removing four holes in the 60--180\textdegree\ range. Fig.~\ref{fig:diversity}(e) demonstrates that this six-hole configuration outperforms the ten-hole design in overall spatial diversity. Therefore, our final design uses a 20~mm diameter cylinder with six strategically placed surface holes.
\squishend

\subsection{Integration of acoustic microstructure to smartphones}
\label{subsec:integration}
\begin{figure}
    \centering
    \includegraphics[width=0.9\linewidth]{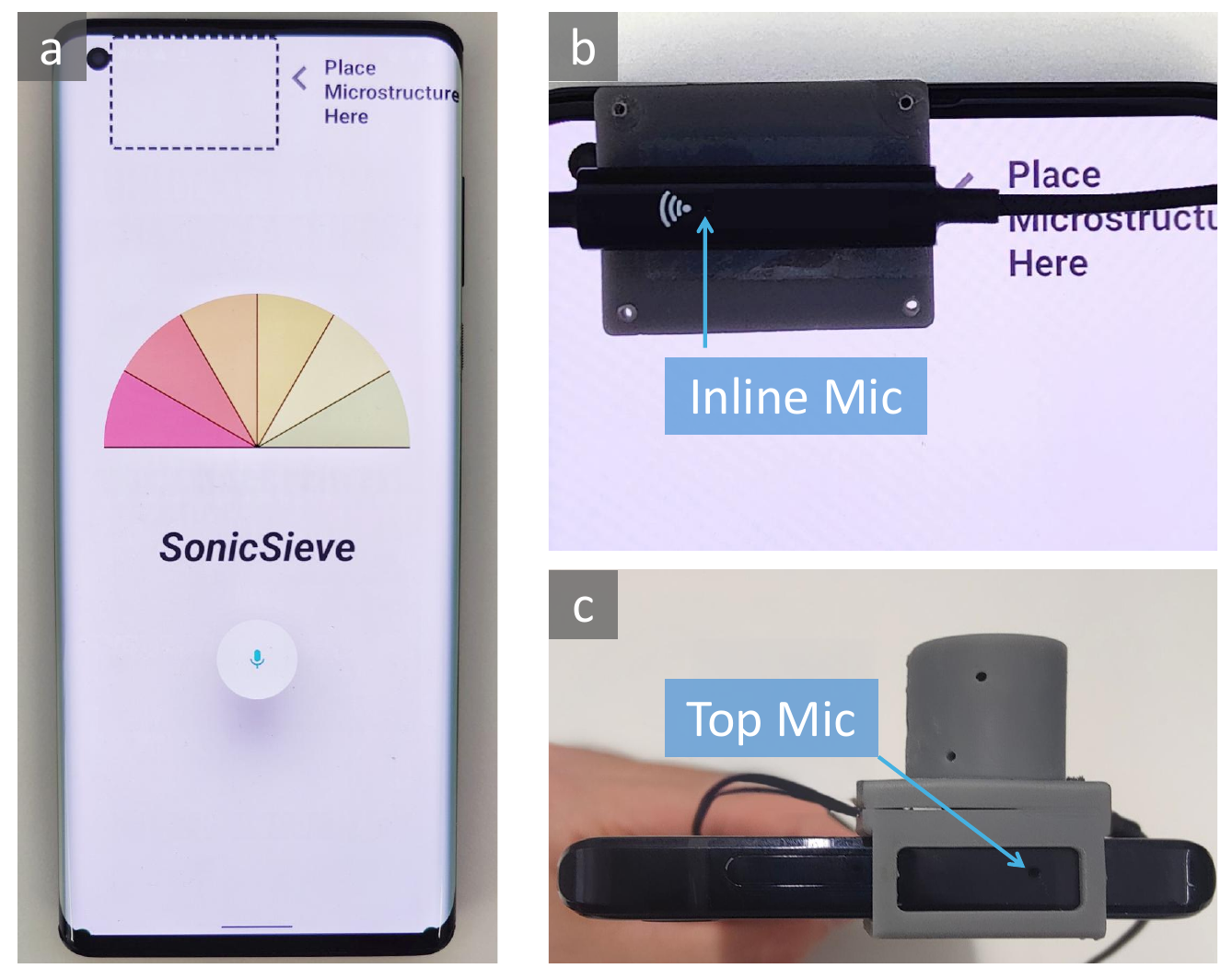}
    \caption{{\bf Illustration of the process to integrate acoustic microstructure to the smartphone.} {\bf (a)} UI on the smartphone app instructs the user to place the microstructure in the proper location. {\bf (b)} Place the inline microphone of a wired earphone on the bottom holder of the microstructure. {\bf (c)} The top part housing the optimized microstructure is attached to the holder using tiny screws. Proper alignment is achieved with the built-in top microphone of the smartphone.}
    \label{fig:assembly}
    \Description{The three sub-figures illustrate the process of integrate acoustic microstructure to the smartphone. The sub-figure (a) is a photo of a smartphone app user interface. The screen displays the name "SonicSieve" at the bottom. A fan-like multicolored graphical element is positioned in the center, and a large circular button is at the bottom. At the top of the screen, there is a text prompt that says, "Place Microstructure Here," with a dashed rectangular box indicating where the microstructure should be placed. The sub-figures (b) and (c) are close-up photographs showing the physical assembly of the microstructure with a smartphone. The sub-figure (b) shows a black plastic holder attached to the bottom of the smartphone. An arrow points to a component labeled "Inline Mic," which is part of a wired earphone. The sub-figure (c) shows the full assembly from a different angle, revealing a top gray plastic part labeled "Top Mic," which houses the optimized microstructure. This top part is attached to the bottom holder using small screws. The figure as a whole visually explains the steps a user would take to correctly attach the acoustic microstructure to a smartphone for the system to function.}
\end{figure}

\noindent \textbf{Microstructure Fabrication.} We implement our optimized microstructure design using computer-aided design and 3D printing. The microstructure geometry is modeled in Rhino 8\footnote{https://www.rhino3d.com/} and fabricated using stereolithography 3D printing. We use ELEGOO Standard Photopolymer Grey Resin\footnote{https://us.elegoo.com/products/elegoo-standard-resin} printed on a Formlabs Form 3 printer\footnote{https://formlabs.com/3d-printers/form-3/}, followed by post-processing with Form Wash for cleaning and Form Cure for curing. This fabrication process ensures precise hole dimensions and smooth internal surfaces critical for acoustic performance.

\noindent \textbf{Assembly Design.} Our integration approach attaches the microstructure to the in-line microphone of standard wired earphones, enabling compatibility with any smartphone equipped with an audio jack. As illustrated in Fig.~\ref{fig:assembly}, the system consists of two main components: a bottom holder that secures the earphone's in-line microphone, and a top housing containing the optimized microstructure. The smartphone app guides users through proper placement, ensuring correct alignment between the microstructure and the device's built-in top microphone.

The final assembly process involves three steps: (1) positioning the earphone's in-line microphone in the bottom holder, (2) attaching the microstructure housing to the holder using precision-fitted connections, and (3) aligning the complete assembly with the smartphone's top microphone. A silicone washer at the interface between components minimizes acoustic leakage and ensures consistent positioning.

\noindent \textbf{Design Benefits.} This integration approach offers two key advantages over device-specific solutions. First, it enables close microphone spacing---our design positions the in-line microphone within 1~cm of the smartphone's built-in microphone, which is critical for minimizing environmental variation effects as established in Section~3.1. Second, the design generalizes across different smartphone models since it relies on standard audio jacks and commonly positioned top microphones rather than custom cases for specific devices.

\noindent \textbf{Microphone Separation Validation.} To validate the importance of close microphone spacing, we empirically measure how separation distance affects signal consistency across environments. Specifically, we measure the signal ratio of two microphones without attaching the microstructure, defined by:
\begin{equation}
    s_{ratio}(f) = \frac{s(f)}{s_{ref}(f)} = \frac{x(f)}{x_{ref}(f)}
\end{equation}
where $x(f)$ and $x_{ref}(f)$ are signals from the inline and reference microphones.

\begin{figure}
\centering
\vspace{-2mm}
\includegraphics[width=0.95\linewidth]
{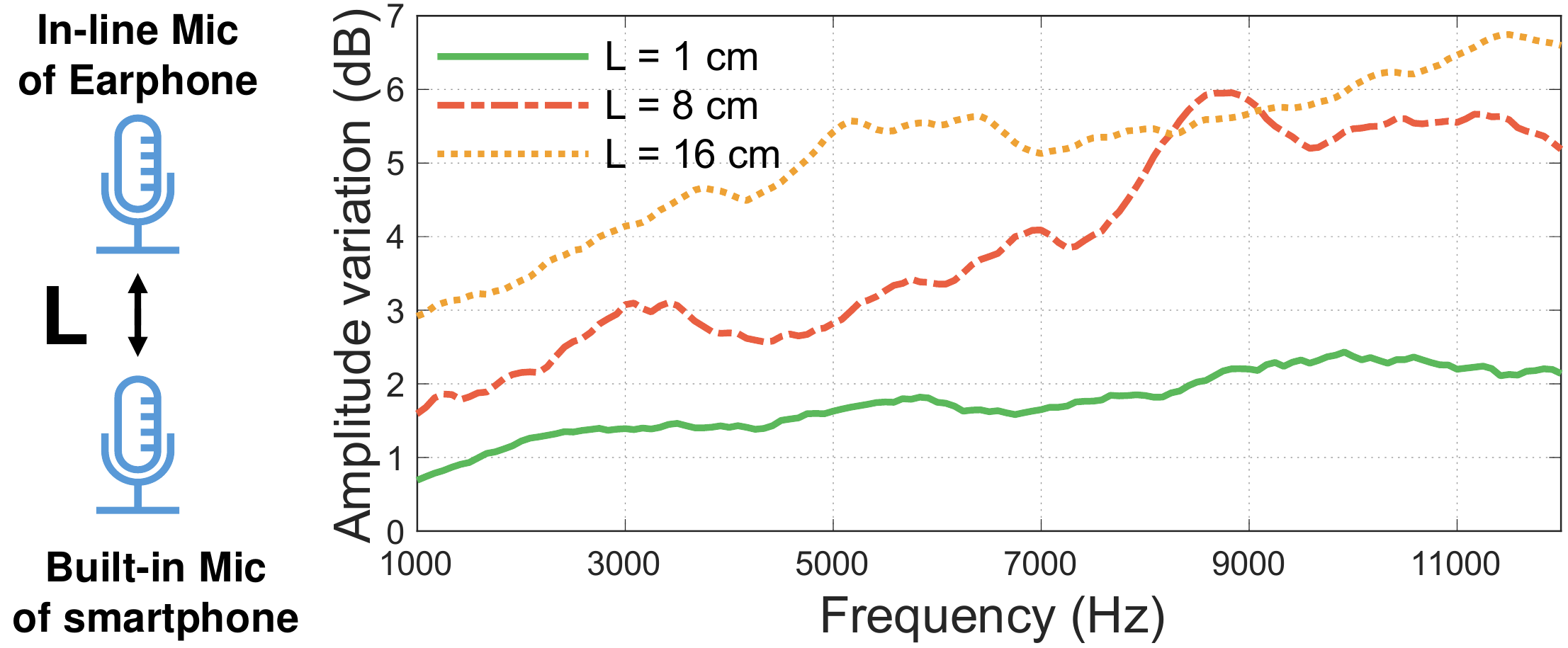}
  \caption{{\bf Effect of separation between earphone's in-line mic and phone's top mic on spectral variation.} We compute the ratio of the signals at the two microphones ($s_{ratio}(f)$) and estimate the variation across environments (lower is better). The variations are lower when the mics are closely spaced.}
  \Description{This figure illustrates the effect of separation distance between earphone's in-line microphone and the smartphone's build-in microphone on spectral variation. The left side shows a diagram of the microphone setup with two blue microphone icons labeled "In-line Mic of Earphone" at the top and "Built-in Mic of Phone" at the bottom, separated by a distance "L" between them. The right side displays a line graph showing "Amplitude Variation (dB)" on the y-axis (ranging approximately from -5 to 15 dB) versus "Frequency (Hz)" on the x-axis ranging from 1000 to 12000 Hz. Three different colored lines represent three separation distances: L = 1 cm (green line at bottom), L = 8 cm (red middle line), and L = 16 cm (orange top line). The graph demonstrates that as the separation distance (L) increases from 1 cm to 16 cm, the amplitude variation between signals received at the two microphones increases significantly. The L = 1 cm condition shows the least spectral variation with a relatively flat response hovering around 0-2 dB across most frequencies. The L = 8 cm shows moderate variation with amplitude ranging from approximately 2-6 dB, with notable peaks and valleys, particularly around 8500 Hz. The L = 16 cm condition shows the greatest amplitude variation across the frequency range, fluctuating between approximately 3-7 dB.}
  \label{fig:mic_separation}
\end{figure}

We conducted experiments across 160 locations in 4 different environments, measuring the magnitude of $s_{ratio}(f)$ at separation distances of 1~cm, 8~cm, and 16~cm. Fig.~\ref{fig:mic_separation} demonstrates that amplitude variation increases dramatically with microphone separation. At 1~cm separation, variations remain below 2.2~dB across all frequencies, while at 16~cm separation, variations exceed 6~dB at higher frequencies. This validates our design requirement for close microphone positioning and confirms that our 1~cm separation distance minimizes environmental artifacts that could interfere with directional cue extraction.



%% file: network-2.tex
\subsection{Directional speech extraction network}

As described in Section~3.1, our microstructure encodes direction-specific spatial cues into the recorded audio signal. However, real-world acoustic environments contain multiple sound sources, environmental reflections, and noise that complicate the extraction of these spatial cues. We develop a deep neural network that processes the two-channel recordings (from the microstructure and reference microphones) to perform directional speech separation in real-time.

\rr{Our network design adapts a state-of-the-art real-time sound separation architecture TF-GridNet~\cite{wang2023tf, chen2024hearable} for angular-based directional speech extraction.} The system processes short audio chunks (8 ms) to enable real-time applications such as live audio streaming and meeting transcription on smartphones.

\noindent \textbf{Pre-processing Module.} Our system processes 2-channel audio input $\mathbf{x} = [x_1,x_2]^T\in \mathbb{R}^{2\times T}$, where $x_1$ and $x_2$ correspond to the reference microphone and microstructure microphone signals respectively, with $T$ representing the audio sample length. To accommodate model lookahead requirements, we apply zero padding to create $\mathbf{x}\in \mathbb{R}^{2\times (T+\sigma)}$, where $\sigma$ is the lookahead sample count.

We implement a learnable Short-Time Fourier Transform (STFT) encoder with sufficiently large window size to enhance model stability across different reverberation environments~\cite{cord2022monaural}. After applying the STFT encoder with discrete Fourier transform length $L$ and step size $\delta$, each microphone channel $i$ produces $X_{i}[f,t]\in \mathbb{R}^{F\times \frac{(T+\sigma-L)}{\delta}}$, where $F$ represents the number of frequency bins.

\begin{figure*}
\centering
\includegraphics[keepaspectratio, height=0.38\textheight]
{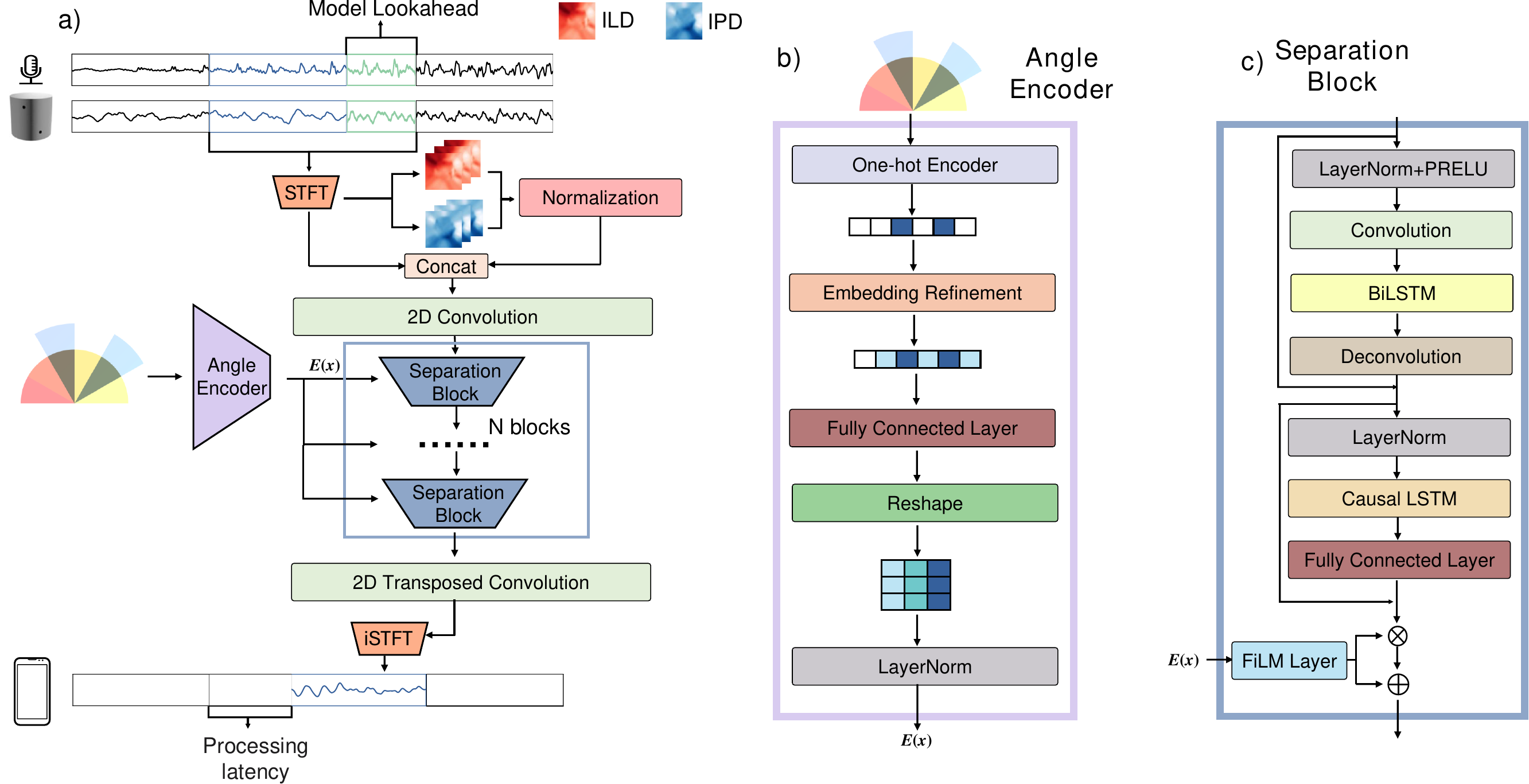}
  \vspace{-0.5em}
  \caption{{\bf Real-time directional speech extraction network architecture.} (a) Audio from the reference mic and the microstructure mic is converted to a spectrogram, ILD/IPD features are extracted, and directional separation is performed based on a user-selected spatial sector. The transformed spectrogram is then converted back to the time domain. (b) The angle encoder converts the selected sector into a normalized embedding for conditioning the separation blocks. (c) Separation blocks use this angle-conditioned embedding to apply directional masking on the spectrogram.}
  \label{fig:network}
  \Description{This figure presents a detailed schematic of a real-time directional speech extraction network architecture divided into three main components (a, b, c): (a) The main processing pipeline shows how audio from two sources (reference mic and in-line mic from microstructure, shown as waveforms at top) is processed through multiple stages: first converting audio to spectrograms, then extracting Interaural Level Difference (ILD) and Interaural Phase Difference (IPD) features, followed by Short-Time Fourier Transform (STFT), concatenation and normalization. The data is processed through a 2D convolution layer, while an angle encoder feeds into multiple separation blocks. Finally, the signal undergoes 2D transposed convolution and inverse STFT (iSTFT), outputting the processed waveform within the time slots after the processing latency. (b) The angle encoder module (detailed in center panel) consists of a one-hot encoder that receives angle information, an embedding refinement layer, a fully connected layer, a reshape operation (visualized as a 3D grid), a layer normalization component, with the output labeled as E(x). (c) The separation block structure (detailed in right panel) contains layer normalization with PReLU activation, convolution layer, Bidirectional LSTM (BLSTM), deconvolution layer, layer normalization, Causal LSTM, fully connected layer, and a final stage with FiLM layer and mathematical operations (multiplication and addition) with the input of E(x).}
\end{figure*}

\noindent \textbf{Feature Extraction.} The core challenge in our approach is extracting the spatial cues embedded by the microstructure while remaining robust to environmental variations and multiple sound sources. While we cannot directly recover the microstructure's direction-specific filter $M_\theta(f)$ due to its entanglement with unknown source signals and environmental effects, we can compute inter-channel difference features that preserve the essential directional information.

We extract interchannel phase differences (IPD) and interchannel level differences (ILD)---well-established spatial cues in binaural hearing research~\cite{blauert1997spatial,gu2024rezero}:

\begin{equation} 
\label{eq:interchannel_features} 
\begin{aligned} 
\text{IPD}[f,t]&=\angle X_2[f,t] - \angle X_1[f,t] \\ 
\text{ILD}[f,t] &= 20 \log \left\{ |X_2[f,t]|/|X_1[f,t]| \right\} 
\end{aligned} 
\end{equation}
The IPD captures phase relationships between the two microphones, which encode timing differences created by the microstructure's directional filtering. The ILD captures magnitude differences that reflect the microstructure's direction-dependent attenuation and amplification patterns. Together, these features encode the spatial signatures that enable directional discrimination.

To facilitate neural network processing, we convert the circular IPD values into continuous representations using trigonometric functions:

\begin{equation} 
X_{feature}[f,t] = \left\{\cos(\text{IPD}[f,t]), \sin(\text{IPD}[f,t]), \text{ILD}[f,t]\right\} 
\end{equation}

We normalize each feature across frequency bands to ensure consistent scaling and improve training stability. For ILD, per-frequency normalization to zero mean and unit variance reduces the impact of frequency-dependent variations inherent in the microstructure design. The cosine and sine IPD components receive similar normalization to balance their influence across frequencies where phase relationships have varying reliability.

The complete feature set combines the real and imaginary STFT components from both microphone channels (4 channels) with our three spatial features, yielding 7 feature channels total. A 2D convolutional block encodes these into the representation $X\in \mathbb{R}^{c\times F\times \frac{(T+\sigma-L)}{\delta}}$, where $c$ is the expanded feature dimension.

\noindent \textbf{Angle Encoder.} We represent user directional selections through a 6-sector encoding system. Selected sectors receive value 1 in a one-hot vector, while adjacent sectors receive 0.25 to provide smooth spatial context. This vector undergoes embedding refinement, fully connected transformations, and reshaping to create directional embeddings $E(\mathbf{x})$ matching our feature dimensions. After LayerNorm, these embeddings condition the separation blocks through Feature-wise Linear Modulation (FiLM).

\noindent \textbf{Real-time Separation Blocks.} Our processing pipeline consists of $N$ sequential LSTM-based blocks optimized for real-time performance. Each block applies LayerNorm and PReLU activation, followed by convolutional downsampling to reduce computational cost for the bidirectional LSTM. After frequency-domain processing, deconvolution with skip connections restores original dimensions.

A causal LSTM with fully connected layer processes temporal relationships while maintaining real-time constraints. The FiLM layer integrates directional information by modulating features according to $E(\mathbf{x})$:

\begin{equation}
E_{sep} = X' \odot \text{Conv1D}(E(\mathbf{x})) + \text{Conv1D}(E(\mathbf{x}))
\end{equation}
where $X'$ represents processed features and $\odot$ denotes element-wise multiplication.

The final output undergoes transposed convolution and inverse STFT to produce time-domain audio. For continuous processing, we output $T$ samples for playback while retaining $\sigma$ samples for subsequent chunk processing, enabling seamless real-time operation.

\begin{figure*}
\centering
\includegraphics[width=.96\linewidth]
{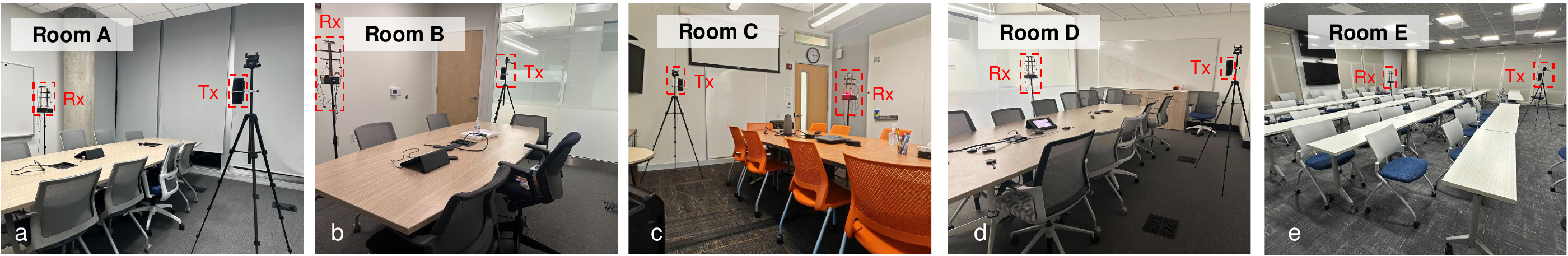}
  \vspace{-1em}
  \caption{{\bf Real-world evaluation setup.} {\bf (a--e)} Five different test environments used for evaluating our system. The room dimensions are as follows (in meters): length $\times$ width $\times$ height. {\bf (a)} Room A, 6.1 $\times$ 4.6 $\times$ 3.7; {\bf (b)} Room B, 4.9 $\times$ 3.6 $\times$ 3.5; {\bf (c)} Room C, 6.1 $\times$ 4.6 $\times$ 3.7; {\bf (d)} Room D, 7.0 $\times$ 4.9 $\times$ 2.7; {\bf (e)} Room E, 11.5 $\times$ 9.1 $\times$ 2.9. In each room we play sound samples from an external speaker (model name: JBL Flip 5), and record it with multiple devices on our data capture rig.}
  \label{fig:realworld}
  \Description{This figure shows a real-world evaluation setup consisting of five different environments (Rooms A-E) used for evaluating the directional speech extraction system. Five sub-figures labeled (a) through (e), each showing a different room setup. Each room contains a data capture rig with recording devices and a tripod setup. Red labels marked as "Tx" and "Rx" are visible in each image, indicating transmitter (speaker) and receiver (data capture rig) positions. The rooms vary significantly in their dimensions and layouts: Room A (a) is a conference room with dimensions 6.1 × 4.6 × 3.7 m, featuring a medium-sized table with chairs; Room B (b) is a smaller meeting room measuring 4.9 × 3.6 × 3.5 m, with a rectangular table and black office chairs; Room C (c) has dimensions 6.1 × 4.6 × 3.7 m, containing orange chairs and table; Room D (d) is a larger conference room measuring 7.0 × 4.9 × 2.7 m, with a long table and multiple chairs; and Room E (e) is the largest space at 11.5 × 9.1 × 2.9 m, showing what appears to be a classroom setting with multiple tables arranged in rows.}
\end{figure*}

\subsection{Real-world dataset}

\noindent \textbf{Motivation and Overview.} Acoustic microstructures encode spatial cues through complex interactions between incoming sound waves and their geometric features. However, these spatial signatures can be significantly altered by environmental acoustic properties such as reverberation, room geometry, and background noise~\cite{wu2016reverberation}. A model trained solely on clean, simulated data may fail to generalize across the diverse acoustic environments encountered in real-world deployments.

To address this challenge, we construct a comprehensive real-world dataset that captures the acoustic variability our system will encounter in practice. Rather than relying on synthetic room impulse responses or simulated acoustics, we physically deploy our hardware in multiple real environments and record actual speech playback. This approach ensures our training data reflects the true complexity of how our microstructure interacts with real acoustic spaces, including reflections, ambient noise, and other environmental factors that synthetic approaches may not accurately model.

Our data collection process involves five distinct rooms with varying acoustic properties, multiple speaker-receiver configurations, and systematic coverage of all directional sectors. This methodology produces a diverse dataset that enables robust performance across different meeting rooms, conference spaces, and classrooms.

\noindent \textbf{Data Sources.} We use speech content from the VCTK~\cite{veaux2017cstr} and LibriTTS~\cite{zen2019libritts} datasets, which provide high-quality recordings of human speech from multiple speakers. All audio clips are resampled to 24 kHz to balance audio quality with real-time processing requirements on mobile devices.

\noindent \textbf{Collection Setup and Procedure.} Our data collection setup consists of a JBL Flip 5 speaker\footnote{https://www.jbl.com/bluetooth-speakers/JBL+FLIP+5-.html} for audio playback and a receiver rig equipped with both a MiniDSP microphone array\footnote{https://www.minidsp.com/products/usb-audio-interface/uma-8-microphone-array} and our \sname\ system (Motorola Edge smartphone with microstructure). The receiver rig includes six LED lasers that divide the surrounding space into 30° sectors, enabling precise speaker placement.

We systematically collect data across five different rooms (meeting rooms, conference rooms, and classrooms) as shown in Fig.~\ref{fig:realworld}. For each room, we place the receiver rig at 3 randomly selected positions. At each receiver position, we place the speaker at 5 different locations distributed across all 6 directional sectors. For each speaker-receiver configuration, we play 6 random 5-second speech clips, resulting in 2,700 individual recordings (6 clips × 6 sectors × 5 speaker positions × 3 receiver positions × 5 rooms).

\noindent \textbf{Mixture Generation and Dataset Statistics.} To create training data for different directional extraction scenarios, we generate synthetic mixtures in software by combining recordings from different sectors. We create three types of mixtures: single-target (6 possible sector selections), two-target (15 combinations), and three-target (20 combinations) scenarios.

For each room and receiver position, this produces 1,200 single-target mixtures, 900 two-target mixtures, and 900 three-target mixtures. We control the Signal-to-Noise Ratio (SNR) between target and interfering sectors within -5 to 5 dB by scaling the audio appropriately. The ground truth for each mixture is the clean combination of target sector signals as recorded by the reference microphone.

Our complete dataset comprises 45,000 mixtures across 5 rooms. We employ leave-one-room-out validation~\cite{wong2015performance} to ensure generalization, reserving one room's data (9,000 mixtures) for testing while using the remaining rooms for training. We additionally reserve 1,200 samples from the training data for validation.

\noindent \textbf{Training Methodology.} We process audio in 8 ms chunks (192 samples at 24 kHz) with 4 ms lookahead for real-time performance. The STFT uses a window size of 288 samples with 192-sample shifts, and we apply 96-sample zero padding at signal boundaries. We optimize using a Scale-Invariant Signal-to-Distortion Ratio (SI-SDR) loss from the Asteroid package~\cite{pariente2020asteroid}:

\begin{align}
\label{eq:loss}
\mathcal{L}_{\text{SI-SDR}} &= 
\begin{cases}
\lambda \| \hat{\mathbf{s}} - \mathbf{s}\|_1 & \text{no speakers in target area} \\
-10 \log \left[ \frac{\|\alpha \cdot \mathbf{s}\|^2}{\|\alpha \cdot \mathbf{s} - \hat{\mathbf{s}}\|^2} \right] & \text{otherwise}
\end{cases}
\end{align}
where $\mathbf{s}$ is the target clean sound, $\hat{\mathbf{s}}$ is the network output, $\lambda = 50$ is a weighting factor, and $\alpha = \frac{\hat{\mathbf{s}}^T \mathbf{s}}{\|\mathbf{s}\|^2}$ provides scale invariance.

We train for 150 epochs using the Adam optimizer~\cite{kingma2014adam}. The learning rate starts at $5\times10^{-4}$, increases to $5\times10^{-3}$ over 10 epochs, remains constant for 20 epochs, then decays by 0.95 every 2 epochs. We apply data augmentation (Time Shift, Time Stretch, Frequency Mask, and Gain Perturbation) with 30\% probability for each method~\cite{Wei2020Augmentation}.

%% file: eval-2.tex
\section{RESULTS}
\subsection{Model results}

\begin{figure*}
        \centering
    \hspace{-1mm}

    \begin{subfigure}{0.3\textwidth}
        \includegraphics[width=\linewidth]{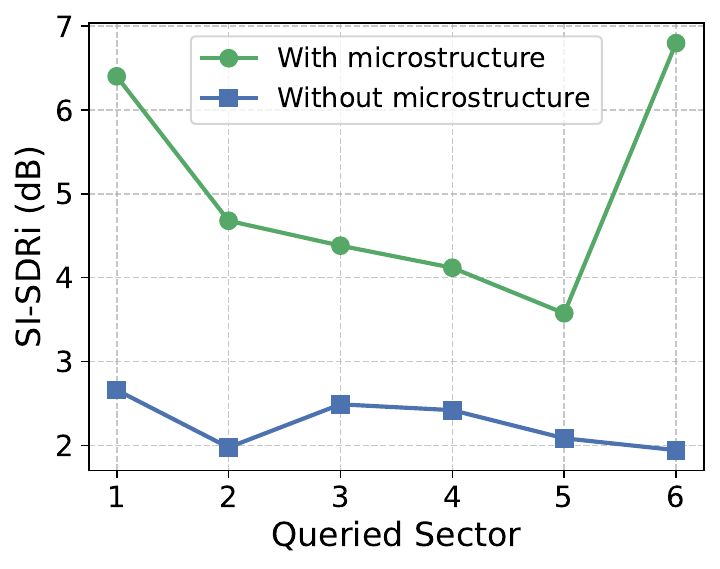}
        \caption{}
        \label{subfig:gain_sectors}
    \end{subfigure}
    \hspace{-1mm}
        \begin{subfigure}{0.3\textwidth}
        \includegraphics[width=\linewidth]{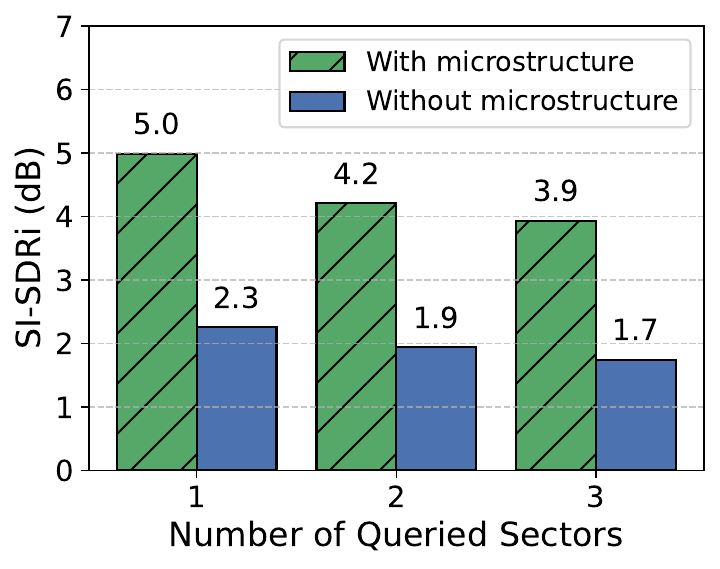}
        \caption{}
        \label{subfig:sector_num}
    \end{subfigure}
    \hspace{-1.5mm}
    \begin{subfigure}{0.395\textwidth}
        \includegraphics[width=\linewidth]{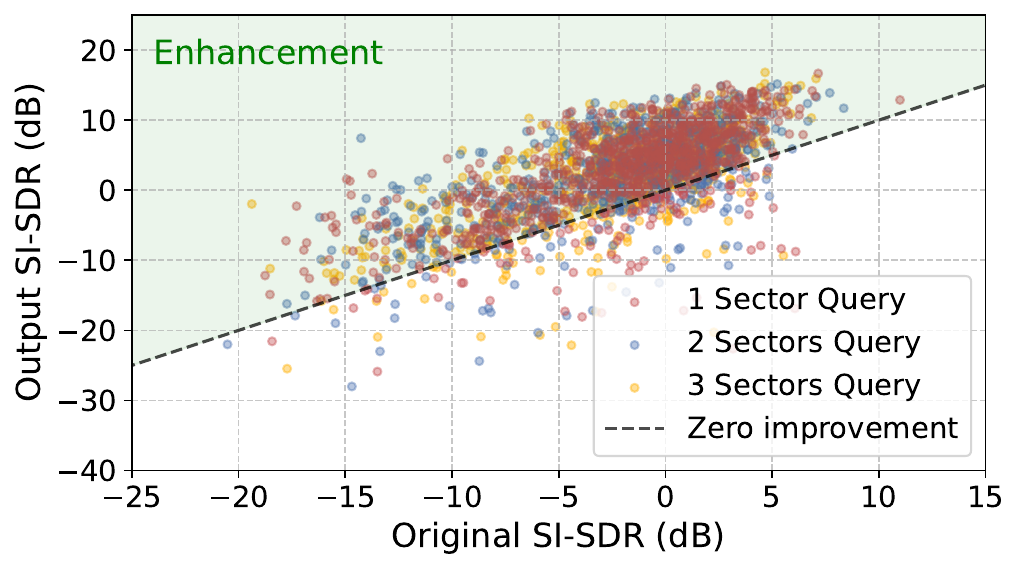}
        \caption{}
        \vspace{2mm}
        \label{subfig:sisdr_scatter}
    \end{subfigure}
    \hspace{-1mm}
    \vspace{-1em}
    \caption{{\bf Overall System performance.  (a)} Effect of selected sector index on SI-SDR improvement.
    {\bf (b)} SI-SDR improvement as a function of the number of simultaneously selected sectors.
    {\bf (c)}  Input vs. output SI-SDR of our model across all samples, grouped by 1-sector, 2-sector, and 3-sector selections. Points above the dashed line indicate cases where SI-SDR improved.} 
    \label{fig:}
    \Description{This figure presents three graphs evaluating system performance through SI-SDR (Scale-Invariant Signal-to-Distortion Ratio) measurements across different conditions: (a) A line graph showing the effect of selected sector index (x-axis, numbered 1-6) on SI-SDR improvement in dB (y-axis, ranging 2-7 dB). Two lines compare performance: With microstructure (light green line with circles): Shows significantly higher SI-SDR values (between 3.6-6.8 dB) with a U-shaped pattern where sectors 1 and 6 perform best. The SI-SDR number for each sector is 6.4, 4.7, 4.4, 4.1, 3.6, 6.8. Without microstructure (dark blue line with squares): Shows consistently lower performance (between 1.9-2.8 dB) with a slight downward trend as sector index increases. The SI-SDR number for each sector is 2.7, 2.0, 2.5, 2.4, 2.1, 1.9. (b) A bar graph comparing SI-SDR improvement (y-axis, in dB) across different numbers of simultaneously queried sectors (x-axis, 1-3 sectors). For each number of sectors, two bars compare: With microstructure (green bars): Shows 5.0 dB for 1 sector, 4.2 dB for 2 sectors, and 3.9 dB for 3 sectors. Without microstructure (blue bars): Shows substantially lower values of 2.3 dB, 1.9 dB, and 1.7 dB respectively.  (c) A scatter plot showing input vs. output SI-SDR (x-axis: Original SI-SDR ranging from -25 to 15 dB; y-axis: Output SI-SDR ranging from -30 to 20 dB). Points are color-coded by query type: 1 Sector Query (red dots), 2 Sectors Query (blue dots), 3 Sectors Query (orange dots). A diagonal dashed line indicates "Zero improvement" reference. Most data points cluster above this line (especially in the -5 to 5 dB original SI-SDR range), demonstrating positive enhancement, with the upper-right quadrant labeled "Enhancement" in green. 
}
\end{figure*}

For our evaluation, we began by using the real-world datasets (Fig.~\ref{fig:realworld}) collected in rooms B-E for training and validation, while testing on data from room A. 
We further present cross-room evaluation results at the end of the section.
In all our evaluations, we use data from the in-line mic version of our system.

\sssec{System performance across different sectors.} We measured the SI-SDR across each of the six sectors in two conditions: before applying the acoustic microstructure and after applying it, with only a single sector selected in both cases. As shown in Fig.~\ref{subfig:gain_sectors}, applying the microstructure consistently achieve higher SI-SDR improvement (SI-SDRi) across all sectors (range: 3.6–6.8~dB; mean: 4.5~dB). Interestingly, we found that performance varies across sectors due to the non-uniform distribution of holes in the microstructure. Specifically, sector 1 and 6 achieve an SI-SDRi of $\sim$ 6-7~dB, while sector 4 and 5 achieve around $\sim$ 4~dB. This variation is consistent with the characterization of the microstructure's spatial diversity for different angles as shown in Fig.~\ref{fig:diversity}, which shows that sectors 1 and 6 correspond to angles with higher spatial diversity, while sectors 3 to 5 exhibit reduced diversity. This characteristic can be leveraged in real-world use where users can rotate the device such that the target speaker is contained by sector 1 or 6 to achieve better directional filtering performance.


\sssec{Effect of number of selected sectors.} As our model is designed to support simultaneous selection of up to three sectors, we conducted a comprehensive evaluation of system performance across all possible sector combinations. This includes evaluating all individual sector (1-sector selection), all possible pairs of sectors (2-sector combinations), and all possible triplets of sectors (3-sector combinations). Each configuration was evaluated both with and without the acoustic microstructure. Our results in Fig.~\ref{subfig:sector_num} reveal two key trends. First, the microstructure-enabled system consistently outperforms the baseline without the microstructure, regardless of selected number. When a single sector is selected, our system achieves a SI-SDRi of 5.0~dB, compared to 2.3~dB for the baseline. With two sectors selected, performance drops slightly to 4.2~dB, but still maintaining an advantage over the baseline of 1.9~dB. When three sectors are selected, the SI-SDRi with the microstructure is 3.9~dB which is greater than the baseline's 1.7~dB. The second trend is that SI-SDRi decreases as the number of selected sectors increases for both systems. This is expected as extracting speech from multiple directions becomes more challenging when the spatial geometry of the selected region becomes more complex.

Fig.~\ref{subfig:sisdr_scatter} shows the relationship between the input SI-SDR of the speech samples in our dataset and the output SI-SDR after processing with our model when 1, 2, or 3 sectors are selected. The diagonal dashed line represents zero improvement, where the output SI-SDR equals the input. Points above this line indicate enhanced speech quality after processing. As evident from the figure, the vast majority of our test samples~(86.6\%) demonstrate positive enhancement, with output SI-SDR values consistently higher than their original values. Notably, the one-sector selection configuration (represented by red dots) shows the fewest samples (9.9\%) falling below the zero-improvement line compared to two- and three-sector selections (15.2\% and 16.3\%, respectively), confirming our earlier observation that the task becomes more challenging as the number of selected sectors increases.

\begin{figure}
    \centering
        \includegraphics[width=0.77\linewidth]{img/hardware_comparison.pdf}
         \vspace{-2mm}
            \caption{ Comparison of SI-SDR improvement across microphone arrays of varying sizes versus our {\sname} system.}
        \label{subfig:num_mics_result}
        \Description{A bar plot compares SI-SDR improvement (y-axis, in dB ranging from 2-5.2 dB) across microphone arrays of varying sizes versus the SonicSieve system. The x-axis shows different configurations: 2-mic: ~2.0 dB, 3-mic: ~3.7 dB, 4-mic: ~4.0 dB, 5-mic: ~4.1 dB, 2-mic (SonicSieve): ~4.5 dB (shown with green diagonal stripes pattern), 6-mic: ~5.1 dB. This demonstrates that the 2-mic SonicSieve system achieves performance comparable to arrays with more microphones, outperforming the 5-mic arrays while using fewer hardware resources.}
    
\end{figure}

\begin{figure*}
        \centering
    \hspace{-1mm}
    
    \begin{subfigure}{0.43\textwidth}
        \includegraphics[width=\linewidth]{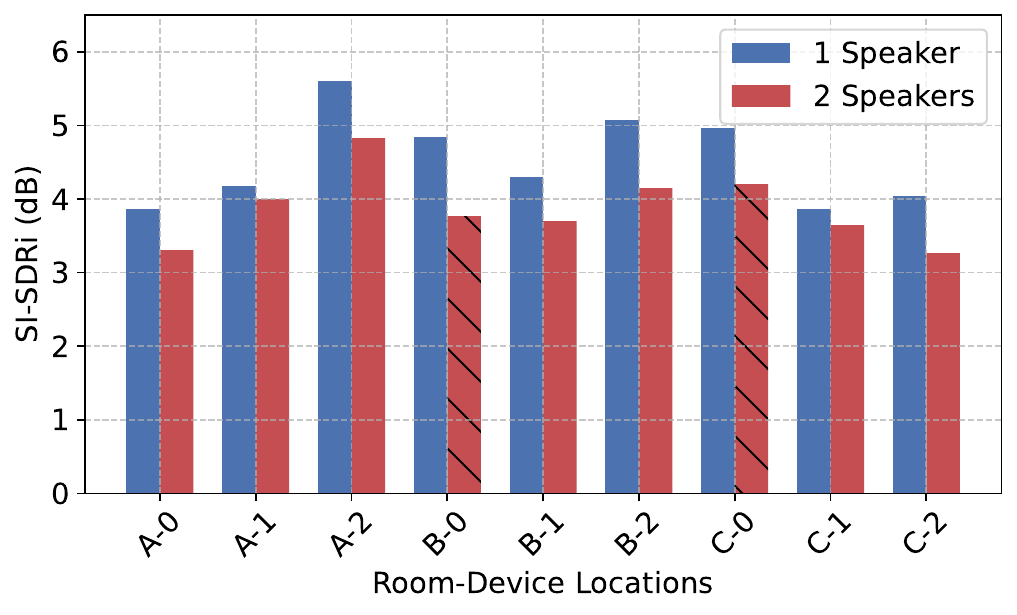}
        \vspace{-5mm}
        \caption{}
        \label{subfig:rooms_generalization}
    \end{subfigure}
    \hspace{-1mm}
    \begin{subfigure}{0.375\textwidth}
        \includegraphics[width=\linewidth]{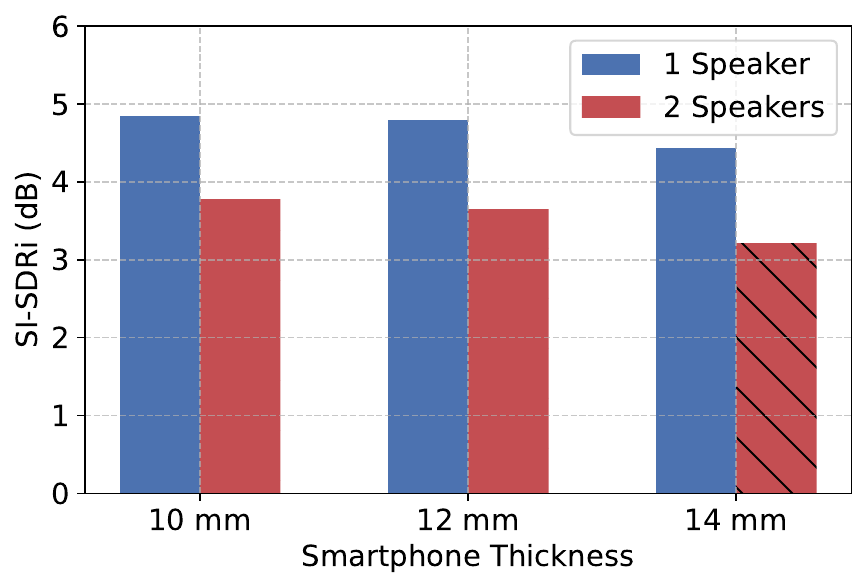}
         \vspace{-3.5mm}
        \caption{}
        \label{subfig:thickness}
    \end{subfigure}
    \vspace{-1em}
    \caption{{\bf Stratified analysis by system variables. (a)} Average SI-SDR improvement across the 3 rooms (Room A, B, C) in our dataset, computed from 3 random locations (location 0, 1, 2) per room. Results are categorized by scenarios with 1 or 2 target speakers and averaged over 3000 samples per location. {\bf (b)} Average SI-SDR improvement with different smartphone thickness. The experiment is conducted at the B-0 location with three prototypes. }
    \label{fig:}
    \Description{ (a) This figure presents a stratified analysis of SI-SDR improvement across different system variables, split into two bar charts:  (a) Right panel shows SI-SDR improvement (y-axis, in dB ranging from 0-6 dB) across 8 different room-device locations (A-0, A-1, A-2, B-0, B-1, B-2, C-0, C-1, C-2) on the x-axis. Each location has two bars: Blue bars: 1 Speaker scenario (values ranging from ~3.8 to ~5.7 dB). Red striped bars: 2 Speakers scenario (values ranging from ~3.2 to ~4.8 dB). In each location, the blue bar is higher than the red bar, meaning that the 1 speaker scenario is simpler than 2 speakers scenario. (b) is a bar chart showing SI-SDR improvement (SI-SDRi) in decibels on the y-axis, ranging from 0 to 6 dB, versus smartphone thickness on the x-axis with three categories: 10 mm, 12 mm, and 14 mm. For each thickness category, there are two bars: a solid blue bar representing "1 Speaker" scenarios and a red hatched bar representing "2 Speakers" scenarios. The blue bars show values of approximately 4.8 dB for 10 mm, 4.8 dB for 12 mm, and 4.4 dB for 14 mm thickness. The red hatched bars show consistently lower values of approximately 3.8 dB for 10 mm, 3.7 dB for 12 mm, and 3.2 dB for 14 mm thickness. Both single-speaker and two-speaker performance show a gradual decline as smartphone thickness increases, with single-speaker scenarios consistently outperforming two-speaker scenarios across all thickness values. A legend in the upper right corner identifies the blue bars as "1 Speaker" and red hatched bars as "2 Speakers."
}
\end{figure*}

\sssec{Comparison against microphone array and traditional beamforming.} We compared the end-to-end performance of \sname\ against a conventional microphone array (UMA-8 USB mic array V2.0) as a baseline. \rr{Our network design is adapted from TF-GridNet~\cite{wang2023tf}, a state-of-the-art real-time sound separation architecture that supports arbitrary numbers of input channels, enabling fair comparison across different microphone configurations.} For this baseline, we used 2 to 6 channels from the array and trained separate neural networks for each configuration. All baseline models were trained on the same dataset with the same architecture as {\sname}, except for minor modifications to accommodate the varying number of input channels.

Fig.\ref{subfig:num_mics_result} shows the SI-SDRi, computed as the difference in SI-SDR between the sound mixture at the reference microphone and the enhanced speech output from the model. As the number of microphone channels increases from 2 to 6, SI-SDRi improves from 2.1~dB to 5.1~dB, due to the increased spatial diversity. \sname\ achieves an SI-SDRi of 4.4~dB, outperforming the 5-microphone configuration (4.1~dB), and approaching the performance of the 6-microphone configuration (5.1~dB). \rr{Notably, the 2-microphone neural baseline achieves only 2.1~dB SI-SDRi, demonstrating that \sname's 4.4~dB improvement stems from the microstructure's ability to fundamentally enrich spatial cues beyond what microphone spacing alone provides.}

\rr{We also evaluated the classical MVDR (Minimum Variance Distortionless Response) beamforming algorithm on the microphone array dataset. MVDR was configured to beamform toward the center of each selected sector (e.g., 15° for sector 1 spanning 0° to 30°). However, MVDR achieved less than 2~dB SI-SDRi even when using all 6 microphone channels. This poor performance stems from several fundamental limitations: (1) the coarse 30° sector resolution cannot provide the precise direction-of-arrival information that MVDR requires for effective spatial filtering, (2) MVDR's linear formulation struggles to handle complex multipath propagation and reverberation, which diffuse the spatial cues, (3) MVDR cannot effectively separate speech from environmental noise when they originate from similar directions, as it relies solely on spatial cues rather than the richer acoustic signatures that neural models can exploit.}



\sssec{Generalization across real-world environments.} The previous results were developed using datasets from Rooms B–E and evaluated in Room A. To further evaluate generalizability of {\sname} across environments, we conducted leave-one-room-out cross-validation for Rooms B and C. Rooms D and E were excluded from this analysis due to data collection inconsistencies in the data collection procedure (with portions of data missing). Fig.~\ref{subfig:rooms_generalization} presents the SI-SDRi performance across various device locations in different rooms, with each label indicating the test room (A, B, or C) and device position (0, 1, or 2) within that room. For single-speaker scenarios (blue bars), our system achieves consistent performance across all test conditions, with SI-SDRi values ranging from 3.9 to 5.7~dB. The highest performance is observed at location A-2 (5.7~dB), while the lowest occurs at C-1 (3.9~dB). Similarly, for two-speaker scenarios (red hatched bars), our system maintains stable performance across all environments, with SI-SDRi values between 3.3 and 4.8~dB. The highest performance is observed at location A-2 (4.8~dB), while the lowest occurs at C-2 (3.3~dB).

\sssec{Impact of smartphone thickness.} \kuang{We evaluated how smartphone thickness affects system performance since our system is designed to generalize to different smartphone geometries. Using our default prototype with the Motorola Edge 2024 (10~mm thickness), we modeled the effects of thicker devices by adding foam tape between the phone screen and microstructure to simulate 12~mm and 14~mm thickness. Fig.~\ref{subfig:thickness} shows that \sname\ maintains robust performance across this thickness range. For single-speaker scenarios, SI-SDRi decreases slightly from 4.84~dB at 10~mm to 4.79~dB at 12~mm and 4.43~dB at 14~mm. Similarly, for two-speaker scenarios, performance drops from 3.77~dB at 10~mm to 3.65~dB at 12~mm and 3.21~dB at 14~mm. While thicker devices show modest performance reduction, the degradation is relatively small (less than 0.6~dB across the tested range). Since most off-the-shelf smartphones have thickness between 8--12~mm, our system demonstrates good compatibility with existing device form factors.}

\begin{figure*}[t]
\centering
\vspace{-2mm}
    \begin{subfigure}{0.28\linewidth}
        \includegraphics[width=\linewidth]{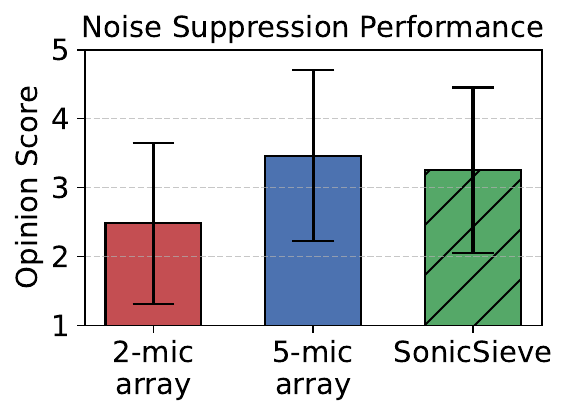}
        \vspace{-5mm}
        \caption{}
        \label{subfig:mos_noise}
    \end{subfigure}
    \begin{subfigure}{0.28\linewidth}
        \includegraphics[width=\linewidth]{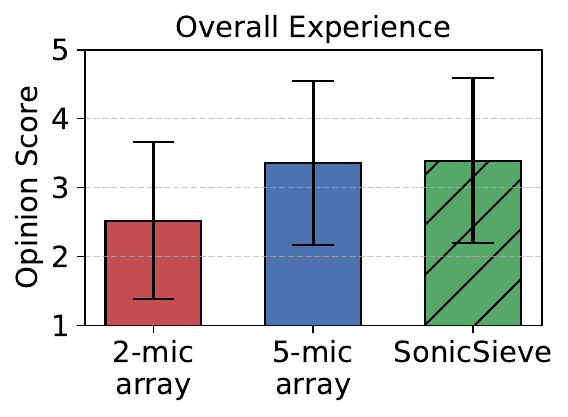}
        \vspace{-5mm}
        \caption{}
        \label{subfig:mos_overall}
    \end{subfigure}
    \begin{subfigure}{0.35\linewidth}
        \includegraphics[width=\linewidth]{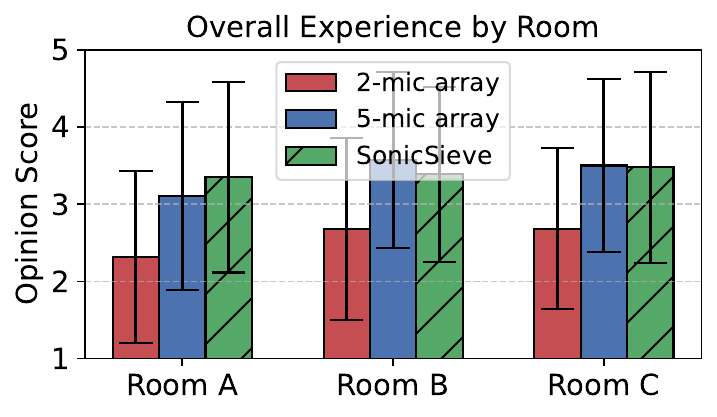}
        \vspace{-5mm}
        \caption{}
        \label{subfig:mos_overall}
    \end{subfigure}
    \vspace{-1em}
  \caption{{\bf User study measuring listening experience via mean opinion scores: (a)} audibility of interfering speakers and background noise, {\bf (b)} overall listening quality, \rr{{\bf (c)} overall listening quality across the audio samples collected in different rooms}}
  \label{fig:mos}
\Description{This figure displays results from a user study measuring listening experience through mean opinion scores on a 1-5 scale (higher is better), presented in three panels with error bars: (a) "Noise Suppression Performance" compares three systems: 2-mic array with mean score of approximately 2.5 (error bars ranging from about 1.3 to 3.7), 5-mic array with mean score of approximately 3.5 (error bars ranging from about 2.1 to 4.9), and SonicSieve with mean score of approximately 3.3 (error bars ranging from about 2.1 to 4.4). (b) "Overall Experience" compares the same three systems: 2-mic array with mean score of approximately 2.5 (error bars ranging from about 1.3 to 3.7), 5-mic array with mean score of approximately 3.3 (error bars ranging from about 2.1 to 4.5), and SonicSieve with mean score of approximately 3.4 (error bars ranging from about 2.1 to 4.7). (c) "Overall Experience by Room" breaks down the overall listening quality across three different rooms (Room A, Room B, Room C), with grouped bars for each system: In Room A, 2-mic array scores approximately 2.2, 5-mic array approximately 3.3, and SonicSieve approximately 3.3. In Room B, 2-mic array scores approximately 2.4, 5-mic array approximately 3.3, and SonicSieve approximately 3.3. In Room C, 2-mic array scores approximately 2.5, 5-mic array approximately 3.5, and SonicSieve approximately 3.5. Panel (a) specifically measures audibility of interfering speakers and background noise (noise suppression capability), while panels (b) and (c) measure overall listening quality. The data shows that SonicSieve and the 5-mic array perform comparably across all metrics and consistently outperform the 2-mic array baseline in all rooms. The overlapping error bars indicate variability in user opinions, but the trends demonstrate clear preferences for the SonicSieve and 5-mic systems across different acoustic environments.}
\end{figure*}

\sssec{Effect of sector resolution.} Although our model is mainly trained in 6 sector cases (30 degree resolution), we conducted evaluation on how the resolution affects model performance. We compared performance between 6 sectors and 9 sectors using simulation in Pyroomacoustics~\cite{scheibler2018pyroomacoustics}. With one speaker in the target sectors, {\sname} achieves 7.4~dB improvement in the 9 sector case and 7.9~dB improvement in the 6 sector case. With two speakers in the target sectors, {\sname} achieves 5.0~dB improvement in the 9 sector case and 5.6~dB improvement in the 6 sector case. This demonstrates that even with more sectors and a more challenging task, {\sname} still achieves good performance.

\sssec{Model Latency.} We evaluated the latency of our directional sound extraction neural network on a variety of different smartphones models: Samsung Galaxy S21, Google Pixel 7, and Motorola Edge 2024. To do this, we exported our PyTorch model to an ONNX model that could be run on these mobile devices. We used an audio chunk size of 8~ms, which we used in our main evaluations. We ran the model for 1000 iterations and obtained mean and standard deviation latencies ranging from 4.5 ± 0.2 to 7.2 ± 0.2~ms across the phone models which is less than the 8~ms requirements needed for real-time execution.

\subsection{Subjective user evaluation of audio quality}
\label{sec:subjective}

While objective metrics like SI-SDR provide valuable quantitative assessment, they cannot fully capture the perceptual quality of speech enhancement systems in real-world scenarios. To evaluate the actual listening experience, we conduct a user study to complement our objective measurements. This study was approved by our Institutional Review Board (STUDY2025\_00000146). All studies complied with relevant ethical regulations. 

\rr{We randomly select 25 generated audio samples from our test set collected from three different rooms (Room A-C).} The selected audio contains one target speech and 1-2 interfering sources (undesired speech or other sound). Each sample is processed by three systems: the 2-channel microphone array (baseline), 5-channel microphone array, and our proposed {\sname}. \rr{44 participants in total (31 males and 13 females) with an average age of 25 years were recruited for the listening test.} The evaluation protocol consisted of three steps: (1) listening to the noisy audio mixture, (2) listening to the clean target speaker audio as reference, and (3) listening to the three processed outputs presented in random order to avoid position bias. Participants were asked to rate each processed audio sample on two criteria using 5-point scales. \kuang{We note that two separate questions~\cite{recommendation2003subjective} are used to decouple the assessment of noise removal from the preservation of target speech quality, as aggressive noise suppression can often introduce undesirable artifacts that a single question would fail to capture.}

\begin{enumerate}
    \item Noise Suppression Performance: \textit{How intrusive/noticeable were the interfering speakers and background noises? (1 -- Very intrusive to 5 -- Not noticeable)}
    \item Overall Experience: \textit{If the goal is to focus on the target speaker, how was your overall experience?" (1 -- Very intrusive to 5 -- Not noticeable)}
\end{enumerate}

As shown in Fig.~\ref{fig:mos}, both {\sname} and the 5-channel microphone array significantly outperform the 2-channel microphone baseline in subjective evaluations. \rr{Across the 25 samples drawn from three rooms, for noise suppression performance~(Fig.~\ref{subfig:mos_noise}), {\sname} achieved a Mean Opinion Score~(MOS) of 3.25 $\pm$ 1.20, compared to the 5-microphone array's 3.46 $\pm$ 1.24 and the 2-channel baseline's 2.48 $\pm$ 1.17. \rr{For overall listening experience} (Fig.~\ref{subfig:mos_overall}), {\sname} achieved \rr{3.39 $\pm$ 1.20}, comparable to the 5-microphone array \rr{(3.35 $\pm$ 1.19)} and higher than the 2-channel baseline (2.52 $\pm$ 1.13). \rr{Fig.~\ref{fig:mos}(c) further breaks down overall experience by room, showing consistent gains over the 2-channel baseline across Room A--C, with {\sname} remaining comparable to the 5-channel array in each environment.}}


\subsection{\rr{User Interface Evaluation}}
\label{sec:ui_eval}

\begin{figure}[h]
\centering
\includegraphics[width=0.75\linewidth]{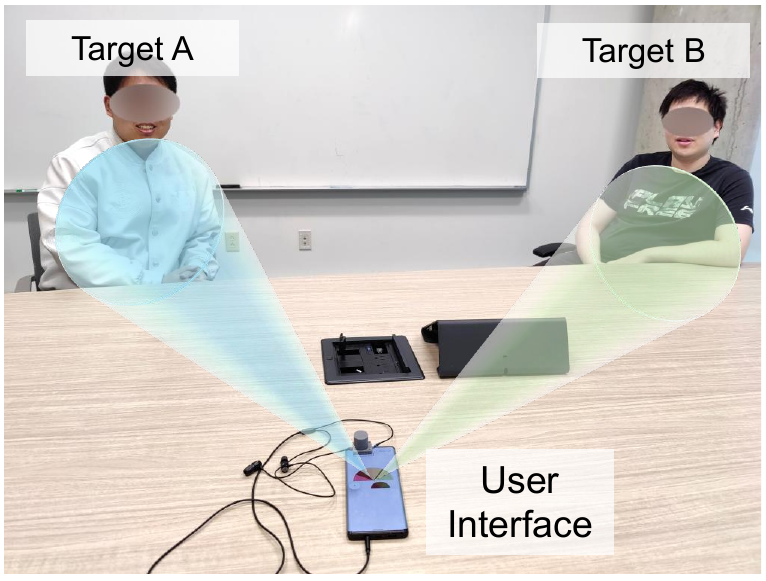}
    \caption{\rr{User interface evaluation setup for directional speech extraction, where the participants are asked to select one or both speakers on the smartphone UI with different configurations in different environments.}}
    \Description{This figure shows a photograph of an experimental setup for evaluating a user interface for directional speech extraction. The scene depicts two participants seated at opposite ends of a table in an indoor environment. Target A, wearing a light blue shirt, is positioned on the left side of the table, while Target B, wearing a green shirt, is positioned on the right side. In the foreground of the image, a smartphone-based user interface device is placed on the table, connected to what appears to be a microphone array with visible wiring. Two semi-transparent colored overlays emanate from the user interface device toward each target speaker: a cyan/blue beam extending toward Target A and a green beam extending toward Target B, visually representing the directional focus capabilities of the system. These colored beams illustrate how the system can selectively focus on individual speakers based on their spatial location. The setup demonstrates the interactive nature of the user interface, where participants can select one or both speakers for audio extraction in different room configurations and acoustic environments.}
    \label{fig:interfaces}
\end{figure}

\begin{table*}[h]
\centering
\large
\caption{\rr{Comparison of system performance and usability across different sector resolutions}}
\label{tab:sector_resolution}
\begin{tabular}{lccccc}
\toprule
\textbf{Config} & \textbf{Resolution} & \textbf{Preference} & \textbf{Accuracy} & \textbf{Confidence (Single)} & \textbf{Confidence (Multiple)} \\
\midrule
\textbf{6 Sectors} & 30° & \textbf{0.96} & \textbf{0.90} & \textbf{0.82} & \textbf{1.00} \\
\textbf{9 Sectors} & 20° & 0.88 & 0.87 & 0.70 & 0.86 \\
\bottomrule
\end{tabular}
\end{table*}

\rr{While our objective metrics demonstrate technical performance, understanding how users interact with directional selection in real-world scenarios is critical for practical deployment. We conduct a user study evaluating the usability of our interface across different sector resolution configurations, examining both selection accuracy and user confidence when targeting speakers in multi-person environments.}

\rr{We recruit 10 participants to evaluate two configurations: our default 6-sector design (30° resolution) and a 9-sector design (20° resolution). As shown in Fig.~\ref{fig:interfaces}, participants were positioned in three conference rooms with two experimenters as target speakers. For each configuration, participants completed three tasks: selecting target A, target B, and both targets simultaneously. Experimenter positions remained consistent across configurations while task order was randomized to minimize learning effects.}

\rr{Participants were instructed to select single or multiple sectors they believed necessary to capture the target speaker(s). We evaluated interface usability through three metrics: (1) \textit{Selection Accuracy}, measured as the probability that selected sectors correctly included the target speakers. (2) \textit{User Confidence}, assessed on a 5-point scale (5=very confident, 1=not confident at all) for both single and multiple speaker selections. (3) \textit{Overall Preference}, rated on a 5-point scale (5=strongly prefer to use, 1=strongly disprefer to use) comparing the two configurations.}

\rr{Table~\ref{tab:sector_resolution} presents our results by normalizing the scores and accuracy to 0--1. The results reveal that the 6-sector configuration consistently outperforms the 9-sector design across all metrics, achieving higher selection accuracy (0.90 vs. 0.87), stronger user preference (0.96 vs. 0.88), and greater confidence for both single-sector (0.82 vs. 0.70) and multiple-sector (1.00 vs. 0.86) scenarios. These findings suggest that finer angular resolution does not necessarily improve usability—instead, the increased cognitive load of managing more sectors can reduce both accuracy and confidence.}

\rr{We observed two key factors driving this preference. First, the coarser 6-sector resolution provides sufficient spatial granularity for typical meeting and lecture scenarios where speakers are distributed around a table or room, while remaining simple enough for rapid selection without requiring precise angular estimation. Second, participants reported that the 9-sector interface felt "overwhelming" and made them second-guess their selections, particularly when attempting to capture multiple speakers. The perfect confidence score (1.00) for multiple-sector selection with 6 sectors suggests that users found it intuitive to select adjacent sectors to create broader capture regions when needed.}

\rr{These results validate our design choice of 6 sectors as providing an effective balance between spatial resolution and interaction simplicity. For future work, adaptive interfaces that automatically adjust sector granularity based on speaker density could potentially combine the benefits of both approaches.}

%% file: discussion-1.tex
\section{LIMITATIONS AND DISCUSSION}


\noindent \rr{\textbf{Dual-microphone design rationale.}} \rr{\sname\ requires two closely spaced microphones for robust directional speech extraction. While the microstructure embeds direction-dependent spectral cues into incoming sound, these cues become entangled with the source's intrinsic spectrum and environmental reflections in single-channel recordings. A dual-microphone design provides a "before/after" comparison---one channel shaped by the microstructure, one unmodified---that allows the neural network to factor out source/room differences and isolate the directional signature. Our choice of an in-line microphone as the second channel offers two key advantages: (1) It can be positioned very close to the phone's built-in top microphone (within 1~cm in our prototype), which our ablations (Fig.~\ref{fig:mic_separation}) show is critical for cross-environment robustness; (2) It is device-agnostic and can work with smartphones that have a standard audio jack or USB-C port. }

\noindent \textbf{Smartphone generalizability.} Our system design prioritizes compatibility across different smartphone models through several key considerations:

\squishlist
\item \textit{Microphone layout adaptation.} Our smartphone application adapts to different device configurations by providing model-specific placement instructions. The UI guides users to position the microstructure optimally based on their device's microphone locations, ensuring proper alignment between the in-line microphone and the smartphone's top microphone (Fig.~\ref{fig:assembly}). 

\item \textit{Generalization across smartphone thicknesses.} Our evaluation across devices ranging from 10 to 14~mm in thickness (Fig.~\ref{subfig:thickness}) shows robust performance with only modest degradation for thicker smartphones. These results indicate that the system can generalize across different form factors. 

\item \textit{Generalization across microphone frequency responses.} To enable generalization across different smartphones microphones, device-specific calibration can be performed by measuring the frequency response of the built-in and in-line microphones, and deriving equalization weights to normalize for their effects. This can allow our system to robustly perform directional speech extraction across a range of devices. A complementary approach is the use of established data-driven approaches such as training on diverse device recordings~\cite{li2017acoustic,ardila2019common} and targeted data augmentation techniques~\cite{morocutti2023device,yuan2025lightweight} which may enable zero-shot generalization without calibration for unseen devices.
\squishend

\begin{figure*}[t]
\centering
\includegraphics[width=.96\linewidth]
{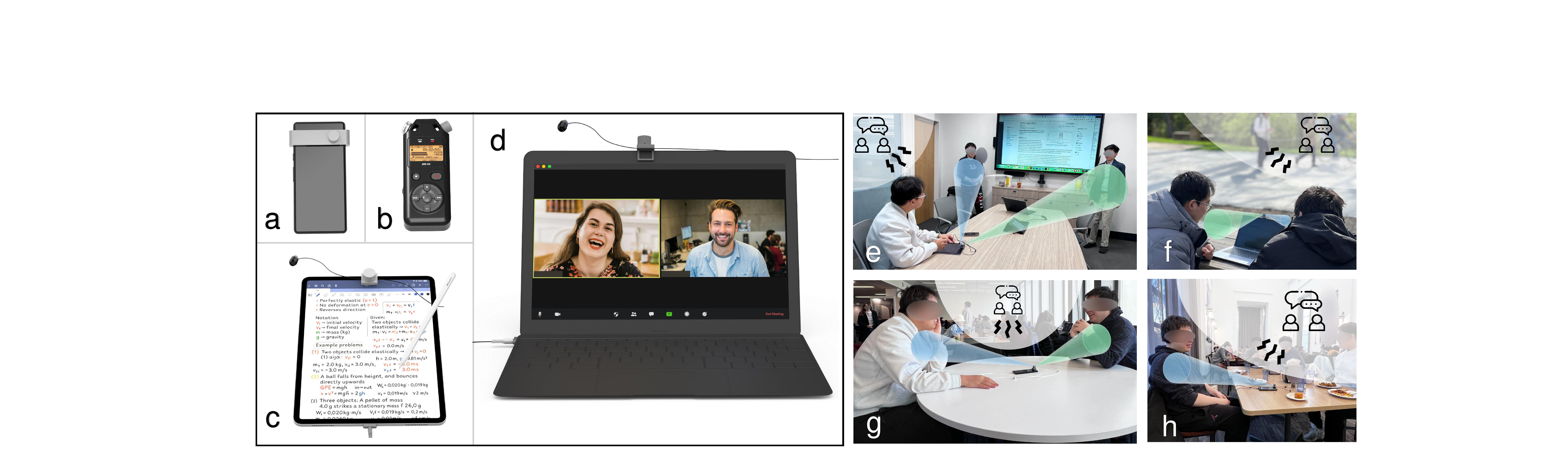}
  \caption{{\bf Applications of {\sname}. (a--d)} Our system can be physically connected to various computing devices with a microphone jack including smartphones, handheld digital voice recorders, tablets, and laptops. {\bf (e--h)} {\sname} can be used in various application scenarios including {\bf (e)} Transcribing a presentation with speaker attribution. {\bf (f)} Conducting a remote meeting by a noisy park. {\bf (g)} Recording a podcast without specialized soundproof booths. {\bf (h)} Interacting with an AI voice assistant in a noisy restaurant.}
  \label{fig:apps}
  \Description{This figure illustrates various applications of SonicSieve across eight labeled panels (a-h), organized in two sections: Left section (a-d) shows physical device compatibility: (a) A smartphone with attached microstructure at the bottom microphone. (b) A handheld digital voice recorder with the microstructure attached at the right top microphone. (c) A tablet with an attached microstructure on the top mic. (d) A laptop showing a video conference call with a microstructure attached on the top mic. Right section (e-h) demonstrates application scenarios with directional audio visualization (shown as green/blue cones indicating audio focus). (e) Two people are presenting a lecture to a student , while the student is using a tablet with microstructure taking notes. (f) Two people conducting a remote meeting in a noisy park using a laptop with a microstructure, with directional focus shown targeting the device user. (g) Two people recording a podcast without specialized soundproof booths at a round table, which can be achieved with a phone with microstructure attached. (h) A person interacting with an AI voice assistant in a noisy restaurant environment. Each application scenario in panels e-h includes small icons of people with speech bubbles and directional indicators, showing how SonicSieve selectively captures audio from specific directions while filtering out surrounding noise. 
}
\end{figure*}


\noindent \rr{\textbf{Pathways to device-native integration.}} \rr{While our current prototype uses a microstructure attached to an external wired earphones, the same core concept can be integrated directly into devices with two closely located microphones. Several integration pathways are feasible: (1) Smartphones with a top and a rear microphone near the camera (e.g., Google Pixel 7) often have $<$2~cm spacing. A microstructure could be embedded near the rear mic while using the top mic as reference (Fig.~\ref{fig:apps}a). (2) Standalone voice recorders with dual front-facing microphones could house the microstructure in a compact form factor (Fig.~\ref{fig:apps}b). (3) Future smartphone designs could incorporate a miniaturized microstructure as part of the device chassis, with one microphone inside the structure and one outside.} 

\noindent {\bf Source separation within a sector.} Our system currently performs spatial filtering along predefined directional sectors and produces a mixture of all sounds it. This can be undesirable when multiple speakers are close to each other and fall within the same sector. Future extensions of this work could address this limitation in two ways: (1) A low-latency speech separation framework~\cite{gu20213d,gu2022towards,yuan2024dewinder} can be applied after directional speech extraction, using permutation-invariant training to separate multiple speakers. (2) More fine-grained separation can be performed in the spatial domain by building on techniques for region-customizable sound extraction~\cite{gu2024rezero} to spatially isolate the target speaker.

\noindent {\bf Speaker mobility.} Our current system is designed primarily for scenarios where the speaker remains relatively stationary such as in meetings. If the speaker moves, the user is expected to manually update the target sector. However, this is not a fundamental limitation of our neural network approach. Future extensions of the model could be explicitly trained to estimate the direction of arrival (DoA) for multiple speakers~\cite{wang2019doa, dey2018direction,Wang2024doa}, and enable dynamic conditioning of the directional speech extraction network.

\noindent {\bf Audio playback.} While our current system is focused on speech transcription and remote streaming, our model's processing latency is low enough for real-time, on-device playback of the extracted audio to users. With a smartphone-native design (Fig.~\ref{fig:apps}a), playback can be achieved using noise-canceling headphones, similar to prior AI-enabled acoustic systems~\cite{veluri2023semantic,chen2024hearable}, which leverage both passive and active noise cancellation to suppress ambient sounds and deliver clear playback of the extracted sounds.

A limitation of the current in-line microphone setup is the close proximity of the mic to the earphones, resulting in insufficient wire length to comfortably reach a user’s ears when the smartphone is placed on a table. This can be addressed in two ways. First, a custom but low-cost headphone could be designed with the in-line mic positioned farther from the speakers, allowing the phone to remain on a surface while maintaining sufficient wire length. Alternatively, wireless headphones could be used for playback by integrating the acoustic microstructure onto one of the outward-facing microphones typically used for noise cancellation. This approach is likely possible as modern headphones often include dual outward-facing microphones for noise cancellation, and low-power GPUs are becoming integrated into wearable devices~\cite{tambe202216,gpuwear}.

\noindent {\bf \rr{Human-centered interaction and deployment considerations.}} \rr{While our user study (Sec.~\ref{sec:subjective} and \ref{sec:ui_eval}) demonstrates the system performance and UI usability in controlled settings, broader human-centered aspects require deeper investigation. Real-world conversational dynamics, including natural turn-taking, situational awareness of non-target speakers, and adaptation to speaker movement, present challenges beyond spatial selection. Accessibility considerations, such as how users with hearing challenges interact with directional audio interfaces, require dedicated study. Similarly, longitudinal in-the-wild deployment across diverse acoustic environments would reveal how interaction patterns evolve over time. Our current work establishes the technical foundation and demonstrates basic usability. These human-centered dimensions represent valuable directions for future research toward broader real-world adoption.}